\documentclass[showpacs]{revtex4}
\usepackage[dvips]{graphicx}
\input{epsf}
\usepackage{amsmath}
\usepackage{amssymb}
\usepackage[mathscr]{eucal}
\usepackage{bm}
\usepackage{theorem}

\newcommand{\beq}{\begin{equation}}
\newcommand{\eeq}{\end{equation}}
\newcommand{\beqa}{\begin{eqnarray}}
\newcommand{\eeqa}{\end{eqnarray}}
\newcommand{\beqan}{\begin{eqnarray*}}
\newcommand{\eeqan}{\end{eqnarray*}}

\newcommand{\de}[2]{ \frac{d #1}{d #2} }

\newcommand{\tr}[1]{{\rm tr} \left( #1 \right) }

\newcommand{\ox}{\otimes}
\newcommand{\op}{\oplus}

\newcommand{\llangle}{\langle\!\langle} 
\newcommand{\rrangle}{\rangle\!\rangle}

\newcommand{\bmrho}{\bm{\varrho}}
\newcommand{\bmlambda}{\bm{\lambda}}

\newcommand{\ad}{{\rm ad}}

\newcommand{\proof}{\noindent {\bf Proof. }}
\newcommand{\qed}{\hfill $\Box$ \vskip 2ex}

\newtheorem{proposition}{Proposition}
\newtheorem{lemma}{Lemma}
\newtheorem{corollary}{Corollary}

{\theorembodyfont{\upshape} 
\newtheorem{remark}{Remark}

}

\begin{document}

\title{Feedback control of spin systems}
\author{Claudio Altafini}
% \thanks{This work was supported by a grant from the Foundation Blanceflor Boncompagni-Ludovisi.}
\affiliation{SISSA-ISAS  \\
International School for Advanced Studies \\
via Beirut 2-4, 34014 Trieste, Italy }
\email{altafini@sissa.it}


\begin{abstract}
  The feedback stabilization problem for ensembles of coupled spin $ 1/2 $ systems is discussed from a control theoretic perspective.
The noninvasive nature of the bulk measurement allows for a fully unitary and deterministic closed loop.
The Lyapunov-based feedback design presented does not require spins that are selectively addressable.
With this method, it is possible to obtain control inputs also for difficult tasks, like suppressing undesired couplings in identical spin systems.
\end{abstract}

\maketitle 

% \begin{keyword}
% matrix Lie groups \sep time-varying differential equations \sep Magnus expansions \sep Wei-Norman formula.
% \end{keyword}

\section{Introduction}

NMR spectroscopy deals with the manipulation of nuclear spins of quantum ensembles, see \cite{Abragam1,Ernst1}.
These systems exhibit most of the essential features of quantum mechanical systems, like the state space of tensorial type (providing exponential growth of the degrees of freedom available) and natural coupling mechanisms between spins, which guarantee the nonclassical nonlocality characteristic of quantum evolutions.
For the purposes of state manipulation, over the last 40 years the field of NMR has developed an extremely versatile and universally accepted set of tools, in the form of sequences of electromagnetic pulses \cite{Ernst1,Cory1,Havel1}. 
In terms of classical control theory, these would be classified as {\em open loop} control methods.
From the perspective of control theory, open-loop methods are by far {\em more complicated} and {\em less robust} than {\em closed-loop} methods, in which one or more functions of the state variables are measured on-line and used to impose a behavoir to the system through a feedback algorithm, or simply to reject errors and disturbances deviating the state from its desired trajectory.
As new directions of research like quantum information processing are pushing for more precise and efficient engineering of quantum states, the use of a quantum/classical interface for the purpose of feedback control of a quantum state is becoming a reasonable task in different settings \cite{Doherty2,Geremia1,Hopkins1,Korotkov1,Steck1,vanHandel1}.
From this control perspective, NMR systems constitute a remarkable opportunity to device feedback control methods at a quantum level for a  number of reasons:
\begin{enumerate}
\item the model of the system is known in detail;
\item its control mechanism is also very accurate;
\item the measurement is classical, thus avoiding all complications due to the state collapse problem (weak or less) unavoidable in other quantum control contexts \cite{vanHandel1,Rabitz1};
\item the relaxation times are sufficiently long to make the real-time interface with a control device feasible.
\end{enumerate}
For all these reasons, a completely classical, unitary and deterministic feedback in the context of NMR systems  is {\em theoretically feasible}, although major technical problems remain to be addressed, like the very low signal-to-noise ratio, the on-line extraction of measurements from coil magnetization in real time and in presence of rf excitation and the need to on-line reconstruct the density matrix from the collective measurements.

The aim of a feedback synthesis is to render a desired state or state trajectory an attractor for the system regardless of the initial condition \footnote{While in open-loop methods the initial condition must always be known, in closed-loop methods no knowledge of the initial state is required.}, and the main purpose of this work is to investigate in detail this feedback synthesis from a theoretical viewpoint.
The main tools we shall use are Lyapunov-based feedback design for bilinear control systems, adapting a method known as Jurjevic-Quinn condition \cite{Jurdjevic5,Bacciotti2} to the case at hand.
Standard references for basic material on feedback control are e.g. the books \cite{Khalil1,Sastry1}.
It is worth mentioning that in the physics literature a related approach (using model-based feedback only for the purposes of attaining the open loop control corresponding to a certain task, see also Sec.~\ref{sec:model-based-feedb}) appears under the name ``tracking control'' \cite{Brown1,Zhu1,Lidar1}.

For the purpose of feedback synthesis, the system is formulated as a bilinear control system living on a compact homogeneous space.
For the task of tracking a given orbit, a class of control Lyapunov functions is naturally defined by the notion of distance induced by the real Euclidean structure with which the homogeneous space is endowed.
This construction resembles closely the Jurdjevic-Quinn stabilization technique \cite{Jurdjevic5} (see also \cite{Ferrante1,Vettori1,Grivopoulos1,Mirrahimi1} for related material dealing with pure states only), although the computation of the largest invariance set via LaSalle principle is more complicated.
This is a consequence of the nontrivial topology of the state space (for a single spin $ 1/2$ it is a sphere $ \mathbb{S}^2 $), implying that no (smooth) feedback design can achieve global stabilization (see \cite{Cla-qu-ens-feeb1} for a detailed analysis in this direction).
At most one can achieve convergence out of a singular set of isolated, repulsive points.
The emphasis on the exact knowledge of the singular locus is motivated by the fact that near a singularity the convergence can be very slow.
Moreover, a local design is of limited practical interest in a quantum context.
For multispin systems it is shown that the tensor product nature of the state space does not complicate exceedingly the feedback synthesis. On the contrary, the singular set of the control law can be computed explicitly thanks to this tensorial structure.

Throughout the paper we consider only the case of spins that are not selectively excitable.
This is clearly the most difficult case, as an rf field affects all spins and interacts with all couplings.
A similar feedback synthesis for selective controls is much simpler (especially for what concerns the convergence analysis) and can be deduced by similar means.

While the analysis is easier for the Ising Hamiltonian, all the results are valid for different types of interactions like Heisenberg or dipole-dipole.
In particular, we show (Sec.~\ref{sec:suppr}) how it is possible to reject unwanted coupling terms, provided they are sufficiently slow compared to the residual nonlocal part of the Hamiltonian.

As an example of a difficult task that can be solved in this way, we consider in Sec.~\ref{sec:model-based-feedb} a system of 3 identical spins in which we want to cancel the interaction between first and third spins without altering the dipole-dipole coupling of the linear chain.
This is a typical problem araising in solid state NMR and for which a pulse sequence is not known.
The feedback scheme allows to compute the time course of the control field achieving an almost exact decoupling in the entire Hilbert space.

\section{Forced Liouville equation for spin $1/2$ systems: a bilinear control model}

The Liouville-von Neumann equation for a density operator $ \rho $ is:
\beq
\dot \rho = - i [ H , \, \rho ] .
\label{eq:Liouville1}
\eeq
For a single spin $ 1/2$ system, assume the Hamiltonian $ H $ is composed of a free part (the drift, often called the Zeeman Hamiltonian) and a forcing part (the control term).
If $  \lambda_j $, $ j=0,\ldots, 3$, are the (normalized: $ \tr{\lambda_j \lambda_k} = \delta_{jk}$) Pauli matrices, the free part, which is due to a strong static magnetic field $ B_o$, is conventionally aligned with the $ \lambda_3 $ axis and causes the spin ensemble to precess around $ \lambda_3 $.
In the laboratory frame this is 
\[
H_{\rm f,\ell}  = - \gamma B_o \lambda_3 
\]
where $ \gamma $ is the gyromagnetic ratio.
The quantity $ \omega_o = \gamma B_o $ is normally referred to as Larmor frequency.
The control Hamiltonian originates from an electromagnetic field rotating in the $ ( \lambda_1, \, \lambda_2 ) $ plane at a frequency $ \omega_{\rm rf} $ close or equal to the precession frequency $ \omega_o$.
In the laboratory frame, this corresponds to the Hamiltonian:
\[
H_{\rm rf} = - \gamma B_1 \left( \cos ( \omega_{\rm rf} t + \phi ) \lambda_1 + \sin ( \omega_{\rm rf} t + \phi ) \lambda_2 \right) 
\]
where $ \phi $ is the phase of the field.
Typically, $ \omega_1 =\gamma B_1 $ is of the order of the tens to hunderds of kHz.
The controllable parameters are the amplitude $B_1 $, the frequency $ \omega_{\rm rf} $ and the phase $ \phi$.
The resulting motion in the laboratory frame is rather complicated to describe.
It is convenient to express it in a rotating frame, i.e., a coordinate system rotating about the vertical axis $ \lambda_3 $ at the frequency $ \omega_{\rm rf}$.
It can be obtained by means of a variation of constants formula.
Up to a negligible term, the neat result is that the inertial Hamiltonian $ H_{\rm f, \ell} + H_{\rm rf} $ is replaced by the rotating frame Hamiltonian $ H_{\rm f} + H_{\rm c} $ given by
\beqa
H_{\rm f} & = & - ( \omega_o - \omega_{\rm rf} ) \lambda_3 =  h^3  \lambda_3 \label{eq:Ham-1spin-rot1} \\
H_{\rm c} & = & - \omega_1 \left( \cos \phi \lambda_1 + \sin \phi \lambda_2 \right) .
\label{eq:Ham-1spin-rot2}
\eeqa
Fixing $ \phi $ means fixing the axis at which the control acts. 
Hereafter we assume $ \phi=0 $.
Hence \eqref{eq:Ham-1spin-rot2} becomes simply:
\beq
H_{\rm c}= u \lambda_1 
\label{eq:Ham-1spin-rot3}
\eeq
with $ u= - \omega_1  $ as our real valued control $ u \in C^\infty (\mathbb{R})$.
When $ \omega_{\rm rf} = \omega_o $, the Hamiltonian is driftless and the resulting motion is called nutation around the axis determined by $ \phi$.

Consider a two spin $ 1/2 $ system in an external field.
To represent the Hamiltonian, we make use of the prodout operators basis given by $ \Lambda_{jk} = \lambda_j \ox \lambda_k $, $ j,k=0,\ldots, 3 $.
Calling $ \gamma_\alpha $ and $ \gamma_\beta $ the gyromagnetic ratios of the two spins, the r.f. Hamiltonian now becomes 
\[
\begin{split}
H_{\rm rf} & = - B_1 \left( \cos ( \omega_{\rm rf} t + \phi )\left( \gamma_\alpha  \Lambda_{10} + \gamma_\beta  \Lambda_{01} \right) \right. \\
&\left.  +  \sin ( \omega_{\rm rf} t + \phi )\left( \gamma_\alpha   \Lambda_{20} + \gamma_\beta \Lambda_{02}  \right) \right)
\end{split}
\]
or, in the rotating frame described above and with $ \phi =0 $
\[
H_{\rm c} = - B_1 \left( \gamma_\alpha  \Lambda_{10}  +  \gamma_\beta  \Lambda_{01} \right).
\]
In the same rotating frame, the free part of the Hamiltonian $ H_{\rm f} $ is 
\beq
H_{\rm f}  = h^{03} \Lambda_{03} +  h^{30} \Lambda_{30} +  h^{33} \Lambda_{33}
\label{eq:Ham-2spin-a} 
\eeq
where $ h^{30}  $ and $ h^{03} $ are the differences between the Larmor frequencies of each spin, call them $ \omega_{o,\alpha} $ and $ \omega_{o,\beta}$, and the carrier frequency $ \omega_{\rm rf} $, $  h^{30}= - ( \omega_{o,\alpha} - \omega_{\rm rf}) $, $ h^{03}= - ( \omega_{o,\beta} - \omega_{\rm rf}) $, and $ h^{33} $ represents the so-called $J$ (or scalar) coupling, with typical values of hundreds of Hz.

If the spins are homonuclear, $ \gamma_\alpha = \gamma_\beta$, then $ \omega_{o,\alpha} $ and $ \omega_{o,\beta}$ (and thus $ h^{03} $ and $ h^{30}$) differ only because of the chemical shift, typically of the order of a few kHz (compared to the MHz order of both $ \omega_{o,\alpha} $ and $ \omega_{o,\beta}$).
If instead we have heteronuclear species, then this difference may be of the order of several MHz.

We shall distinguish between nonselective and selective Hamiltonians.
The main difference between the two cases is in the structure of the control Hamiltonian $ H_{\rm c} $.
In the first case, it is assumed that the different spins all experience the effect of the same control field
\beq
H_{\rm c}  =  u H_{\rm c, ns}= u \left(  \Lambda_{01} +  \Lambda_{10} \right) 
\label{eq:Ham-2spin-b}
\eeq
where $ u = - \gamma_\alpha B_1  =- \gamma_\beta B_1 $.
This model is suited for homonuclear species with $ \omega_{o, \alpha} $ and $  \omega_{o,\beta} $ not sufficiently separated.
When instead $ \omega_{o, \alpha} $ and $  \omega_{o,\beta} $ are well separated, like in homonuclear species with consistent chemical shift or in heteronuclear species, it is a good approximation to consider the spins as selectively excitable \footnote{Selectivity normally holds when $ | \omega_{o, \alpha} - \omega_{o,\beta} | > | \omega_1 | $. In a feedback design, $ | \omega_1 | \simeq k $, the feedback gain. Therefore high gain feedback may reveal unsuitable for selective controls, especially in the homonuclear case.}.
This is achieved by considering 2 different rf fields tuned around the two frequencies $ \omega_{o,\alpha} $ and $ \omega_{o, \beta} $ call them $ \omega_{{\rm rf},\alpha} $ and $ \omega_{{\rm rf}, \beta} $.
The separation of $ \omega_{o,\alpha} $ and $ \omega_{o, \beta} $ implies that the cross-talk with the off-resonant spin is negligible.
By applying the double coordinate change, one gets still the free Hamiltonian \eqref{eq:Ham-2spin-a}, but now with $ h^{30} = - (  \omega_{o,\alpha} - \omega_{{\rm rf},\alpha})  $ and $ h^{03} = - (  \omega_{o,\beta} - \omega_{{\rm rf},\beta})  $.
Choosing $ \omega_{{\rm rf},\alpha} =  \omega_{o,\alpha} $ and $  \omega_{{\rm rf},\beta} =  \omega_{o,\beta}$, the two local precessions disappear from \eqref{eq:Ham-2spin-a}.
The coupling term is unchanged, as $ \Lambda_{33} $ commutes with $ \Lambda_{03} $ and $ \Lambda_{30}$.
The control Hamiltonian, instead, becomes
\beq
H_{\rm c}  =  u_{01}  \Lambda_{01} + u_{10}  \Lambda_{10}  
\label{eq:Ham-2spin-c}
\eeq
where $ u_{01} = - \gamma_\beta B_{1, \beta}  $ and $ u_{10} = - \gamma_\alpha B_{1, \alpha}  $.

Models like \eqref{eq:Ham-2spin-a} with only a vertical coupling are often referred to as Ising Hamiltonians.
In particular, notice that since $ H_{\rm f} $ is diagonal,
\[
H_{\rm f} = \begin{bmatrix} 
 h^{03} +  h^{30} +  h^{33}\! \!\! \!\! \!\! \!  & & & \\
&\! \!\! \!\! \!\! \!\! \!\! \! - h^{03} +  h^{30} -  h^{33} \! \!\! \!\! \!\! \!\! \!\! \!& & \\
& &\! \!\! \!\! \!\! \!\! \!\! \! h^{03} -  h^{30} -  h^{33} \! \!\! \!\! \!\! \!\! \!\! \! & \\
& & &\! \!\! \!\! \!\! \!\! \!\! \! - h^{03} -  h^{30} +  h^{33}  
\end{bmatrix},
\]
when $ h^{03} = h^{30} $ \footnote{Since $ H_{\rm f} $ is relative to at least a carrier frequency (two in the selective case \eqref{eq:Ham-2spin-c}), it is not a restriction to assume $ h^{03} $, $ h^{30}  \geqslant 0$} the unforced system has at least one degenerate eigenvalue of multiplicity 2, regardless of the value of $ h^{33} $.
This may result in a loss of controllability and complicates also the convergence in the closed-loop system.
The diagonal elements of $ H_{\rm f} $ have the meaning of energy levels of the (unperturbed) system.
From \cite{Cla-contr-root1}, since $ {\rm Graph}(H_{\rm c,ns} )$ is connected, as soon as $ H_{\rm f} $ is $ H_{\rm c,ns} $-strongly regular, i.e., has energy levels that are nondegenerate and transition frequencies all different in correspondence to the nonzero elements of $ H_{\rm c,ns} $, then the system is controllable, see Theorem~3 of \cite{Cla-contr-root1}.

\begin{lemma}Consider the system \eqref{eq:Ham-2spin-a}-\eqref{eq:Ham-2spin-b}.
$ H_{\rm f} $ is $ H_{\rm c,ns} $-strongly regular if $ h^{03} \neq h^{30} $, $ h^{33} \neq 0 $ and $h^{33} \neq \pm ( h^{03} - h^{30})/2$.
\end{lemma}
\proof
The graph of the control Hamiltonian $  H_{\rm c,ns} $ habilitates the following 4 (nonoriented) transitions: $ 1\leftrightarrow 2$, $ 1\leftrightarrow 3 $, $ 2\leftrightarrow 4$, $ 3\leftrightarrow 4 $.
Computing the energy differences in terms of the $ h^{03}$, $ h^{30} $ and $ h^{33}$, lack of degenerate transitions corresponds to the three inequalities stated above.
\qed

The condition of the Lemma is a generic condition, satisfied almost always.
Furthermore, it is sufficient but not necessary for controllability.
\begin{corollary}
The system \eqref{eq:Ham-2spin-a}-\eqref{eq:Ham-2spin-b} is controllable if $ h^{03} \neq h^{30} $, $ h^{33} \neq 0 $ and $h^{33} \neq \pm ( h^{03} - h^{30})/2$.
\end{corollary}

For $ n$ spin $ 1/2$, in a rotating frame of frequency $\omega_{\rm rf} $, the Ising Hamiltonian of a linear spin chain is still composed of a drift part containing the Larmor precessions (relative to $ \omega_{\rm rf} $ as in \eqref{eq:Ham-2spin-a}) plus the $J$ couplings between adjacent spins
\[
\begin{split}
H_{\rm f} = &   \left( h^{0\ldots 03 }\Lambda_{0\ldots 03}  + \ldots +  h^{30\ldots 0 }\Lambda_{30\ldots 0} \right.\\
&\left. +  h^{0\ldots 033 }\Lambda_{0\ldots 033}  + \ldots +  h^{330\ldots 0 }\Lambda_{330\ldots 0} \right) ,\end{split}
\]
and of a forcing term along the $\lambda_1 $ axis of each spin which, in the nonselective case, is
\[
H_{\rm c}  =u H_{\rm c,ns}=  u \left(\Lambda_{0\ldots 01}  + \ldots + \Lambda_{10\ldots 0}\right) .
\]

Rather than working with the complex matrix ODE \eqref{eq:Liouville1}, (whose integral flow corresponds to a conjugation action) we prefer to use the real ``one-sided'' linear action on vectors (or tensors) of expectation values along a complete orthonormal set of observables.
For a single spin $ 1/2 $, it is well-known that one can write the density operator $ \rho $ as a real vector of expectation values along the Pauli matrices $ \lambda_j $, $ j=0,\ldots, 3$: $ \rho= \varrho^j \lambda_j =  \bmrho \cdot \bmlambda $, using the summation convention.
In terms of $\bmrho $, and for the Hamiltonian in \eqref{eq:Ham-1spin-rot1} and \eqref{eq:Ham-1spin-rot2} one obtains the Bloch equations:
\beq
\dot \bmrho = -i \left( h^3 \ad_{\lambda_3} + u \ad_{\lambda_1} \right) \bmrho .
\label{eq:contr-liouv-1spin}
\eeq
The notation ``ad'' in \eqref{eq:contr-liouv-1spin} originates from the notion of adjoint representation, and the adjoint operators $ {\rm ad}_{\lambda_j}  $ stand for matrices of structure constants with respect to the $ \mathfrak{su}(2) $ basis given by the $ -i \lambda_j $: $ {\rm ad}_{\lambda_j} \lambda_k = [ \lambda_j, \, \lambda_k ] = \sum_{l=0}^3 c_{jk}^l \lambda_l $. 
In general, the adjoint representation of a semisimple Lie algebra is a real isomorphic matrix representation of the algebra.
For $ \mathfrak{su}(2) $ for example, we have $ \ad_{\mathfrak{su}(2)} = \mathfrak{so}(3)$.
The most important feature of the adjoint representation in our case is that it provides a linear representation of one-parameter groups of automorphisms of the algebra.
This enables us to formulate the control problem in terms of standard real bilinear control systems also for multispin systems.
In fact, for 2 or more spin $ 1/2 $ densities, a parametrization similar to the Bloch vector yields a tensor, called the Stokes tensor and also denoted by $ \bmrho$: if $ \rho \in {\cal H}_2 ^{\ox 2 }$, $ \rho = \varrho^{jk} \Lambda_{jk}= {\bmrho \cdot {\bm \Lambda}} $ where $ \varrho^{jk} = \tr{\rho \Lambda_{jk} } $ are expectation values capturing all 15 degrees of freedom of $ \rho $ along the complete set of observables $ {\bm \Lambda} = \{  \Lambda_{jk}, \; j,k=0,\ldots, 3 \} $, $ \bmrho \in {\cal S} \subsetneq \mathbb{S}^{15}$, where the real set $ {\cal S} $ is very difficult to describe explicitly.
All details are given in \cite{Cla-spin-tens1}.
Calling 
\[
{\rm ad}_{\Lambda_{jk}}  =  \frac{1}{2} \left( {\rm ad}_{\lambda_j } \otimes  {\rm aad}_{\lambda_k } +  {\rm aad}_{\lambda_j } \otimes  {\rm ad}_{\lambda_k } \right) 
\label{eq:adLjk}
\]
the real skew-symmetric operators obtained from the $ \ad_{\lambda_j} $ above and the ``antiadjoint'' operators $ {\rm aad}_{\lambda_j} $ (which have a similar meaning, only involving the ``symmetric'' structure constants $
{\rm aad}_{\lambda_j } \lambda_k = \{ \lambda_j , \, \lambda_k \} = \sum_{l=0}^3 s_{jk}^l \lambda_l $),
then we obtain the following adjoint representation of the Liouville equation
\beq
\dot{\bmrho} = -i \left( \ad_{H_{\rm f}} + u \ad_{H_{\rm c,ns}}  \right) \bmrho  
\label{eq:liov-2spin}
\eeq
or, in components,
\beq
\begin{split}
\dot{\rho}^{pq}  = & 
 - i \left( h^{03}  {\rm ad}_{\Lambda_{03}} +  h^{30}  {\rm ad}_{\Lambda_{30}} +  h^{33}  {\rm ad}_{\Lambda_{33}}  \right) _{lm}^{pq} \rho^{lm} \\
& 
- i   u \left(   {\rm ad}_{\Lambda_{01}} +  {\rm ad}_{\Lambda_{10}}  \right) _{lm}^{pq} \rho^{lm}  .
\end{split}
\label{eq:liov-2spin-comp}
\eeq
By writing $ \rho^{jk} $ as a 16-vector and expanding the tensor products, a bilinear control system with drift and control vector fields that are $ 16 \times 16 $ matrices is obtained.
The expression is similar for the selective case \eqref{eq:Ham-2spin-a}-\eqref{eq:Ham-2spin-c}.

Obviously, from $ \mathfrak{g}_{2s} = {\rm Lie} \{ -i \Lambda_{jk}, \, j,k =0,\ldots, 3 \} = \mathfrak{su}(2)\op  \mathfrak{su}(2) \cup \mathfrak{su}(2)\ox  \mathfrak{su}(2) $, one gets for the adjoint representation $\ad_{\mathfrak{g}_{2s} }= {\rm Lie} \{ -i \ad_{\Lambda_{jk}}, \, j,k =0,\ldots, 3 \} = \mathfrak{so}(3)\op  \mathfrak{so}(3) \cup \mathfrak{so}(3)\ox  \mathfrak{so}(3) $.

In the case $ u=0 $ the Ising model is completely integrable because 
\beq 
[ \ad_{\Lambda_{03}}, \ad_{\Lambda_{30}} ] = [ \ad_{\Lambda_{03}}, \ad_{\Lambda_{33}} ] = [ \ad_{\Lambda_{30}}, \ad_{\Lambda_{33}} ] = 0.
\label{eq:comm-drift-2spin}
\eeq
Hence 
\beq
\varrho^{pq} (t) = \left( {\rm exp} \left( -i t h^{03} \ad_{\Lambda_{03}} \right)  {\rm exp} \left( -i t h^{30} \ad_{\Lambda_{30}} \right)  {\rm exp} \left( -i t h^{33} \ad_{\Lambda_{33}} \right) \right)_{lm}^{pq}  \varrho^{lm} (0) 
\label{eq:int-exp-drift-2spin}
\eeq
which shows that the evolution is a ``superposition'' of 3 different periodicities, two local $ \tau_{p_\alpha} = \frac{2 \pi}{h^{30}} $ and $ \tau_{p_\beta} = \frac{2 \pi}{h^{03}} $ and one nonlocal $ \tau_p= \frac{2 \pi}{h^{33}} $.
The purity of the two reduced densities $  \rho_\alpha (t) = {\rm tr}_\beta (\rho )  $ and $  \rho_\beta (t) = {\rm tr}_\alpha (\rho ) $ is varying with time because of the coupling.

The generalization to $ n$ spin $ 1/2$ is completely analogous: the $ 2^n \times 2^n $ density matrix $ \rho $ can be described by an $n$-index tensor $ \rho = \varrho^{j_1 \ldots j_n } \Lambda_{j_1 \ldots j_n } =\bmrho \cdot {\bm \Lambda } $, each index ranging in $ 0, \ldots, 3 $, $ \Lambda_{j_1 \ldots j_n } = \lambda_{j_1} \ox \ldots \ox \lambda_{j_n} $.
The corresponding ODE is still given by an equation like \eqref{eq:liov-2spin}.

%%%%%%%%%%%%%%%%%%%%%%%%%%%%%%%%%%%%%%%%%%%%%%%%%%%%%%%%%%%%%%%%%%%%%%%%%%%%%%%%
\section{State feedback stabilization of spin-$1/2$ systems}

In this Section, we are only interested in full state feedback.
The topology of the manifolds discussed in this work (spheres and compact homogeneous spaces obtained by taking the ``envelope'' of tensor products of ``affine'' spheres) forbids to have globally converging smooth algorithms. For example, for the Bloch sphere $ \mathbb{S}^2 $ there does not exist smooth positive definite functions with less than two points having zero derivative.
In control theory, the functions having such minimal number of zeros are sometimes referred to as Morse functions \cite{Koditschek3}.
Hence in the simplest case of a single spin $ 1/2 $, the Lyapunov-based design will always be characterized by the presence of at least a spurious equilibrium point, which can, however, be rendered repulsive.

\subsection{State feedback stabilization for a single spin-$1/2$ system}
\label{sec:one-spin-feedb-laws}

Because of the drift term, for $ \bmrho $ different from the north/south poles of $ \mathbb{S}^2$ the unforced system $ \dot \bmrho = -i h_3 \ad_{\lambda_3} \bmrho  $ does not have an equilibrium point but keeps moving on an orbit in the $ (\varrho_1,\varrho_2) $ plane corresponding to $ \varrho^3 (t) = \varrho^3 (0)$.
We shall not try to cancel such precession motion, but instead try to stabilize the state to a desired orbit.
Consider the system \eqref{eq:contr-liouv-1spin} from a given initial condition $ \bmrho(0) $.
If $ \bmrho_{\rm d} $ is the reference state, describe the orbit to track (with the obvious prerequisite $ \| \bmrho_{\rm d}\| = \| \bmrho \|= r $) by means of a ODE like \eqref{eq:contr-liouv-1spin} but without forcing terms.
Calling $ \bmrho_{\rm d}(t)$ the desired reference state, 
\beq
\dot \bmrho_{\rm d} = -i  h^3_{\rm d} \ad_{\lambda_3} \bmrho_{\rm d} 
\label{eq:liouv-1spin-d}
\eeq
means that $ \varrho^1_{\rm d} $ and $ \varrho^2_{\rm d}$ evolve on a circle while $ \varrho^3_{\rm d}(t) =  \varrho^3_{\rm d}(0) $ is the fixed value that characterizes the orbit.

\begin{proposition}
\label{prop:orb-tr-1spin-1c}
The system \eqref{eq:contr-liouv-1spin} with the feedback law 
\beq
u = k \llangle \bmrho_{\rm d}, \, -i  \ad_{\lambda_1} \bmrho \rrangle,
\label{eq:feedb-cotr-1-spin2-1c}
\eeq
where $ k \in \mathbb{R}^+$, is tracking the reference orbit $ \bmrho_{\rm d}(t)$ given by \eqref{eq:liouv-1spin-d} with $ h^3_{\rm d} = h^3 $, in an asymptotically stable manner, for all $\bmrho(0) \in \mathbb{S}^2_r  $ with the exception of $ \bmrho(0) = - \bmrho_{\rm d} (0)$ and of $ \bmrho(0) $, $ \bmrho_{\rm d} (0) $ such that $ \varrho^3 (0) = \varrho^3_{\rm d}=0 $.
\end{proposition}

\proof

For density operators, a natural choice of distance is given by the so-called Hilbert-Schmidt (or trace) norm: $ \tr{\rho^2}$. 
In terms of Bloch vectors $ \bmrho_{\rm d} $, $ \bmrho$, this induces the following $ \mathbb{S}^2 $ distance between $ \bmrho_{\rm d} $ and $ \bmrho $, see e.g. \cite{Zyczkowski2}:
\beq
d(\bmrho_{\rm d} , \bmrho ) = \| \bmrho_{\rm d} \| ^2  - \llangle \bmrho_{\rm d}, \bmrho \rrangle 
= \| \bmrho_{\rm d} \| ^2  - \bmrho_{\rm d}^T  \bmrho .
\label{eq:dist-1spin}
\eeq
Consider as candidate Lyapunov function the distance \eqref{eq:dist-1spin}: $ V  = d(\bmrho_{\rm d} , \bmrho )  $.
Clearly $ V\geqslant 0$ and $  V =0 $ only when $ \bmrho_{\rm d} = \bmrho $.
Since $ \de{\|\bmrho_{\rm d}\|^2}{t} =0 $
\beq
\begin{split}
\dot V  & =  - \llangle \dot \bmrho_{\rm d} , \bmrho \rrangle -  \llangle \bmrho_{\rm d} , \dot \bmrho \rrangle \\
& =  - \llangle -i  h^3 \ad_{\lambda_3} \bmrho_{\rm d} , \bmrho \rrangle -  \llangle \bmrho_{\rm d} , -i \left(  h^3 \ad_{\lambda_3} + u \ad_{\lambda_1} \right)   \bmrho \rrangle \\
& = i u \bmrho_{\rm d} ^T \ad_{\lambda_1} \bmrho ,
\end{split}
\label{eq:dotV-1spin-1c}
\eeq
because $ i\ad_{\lambda_3} $ is skew-symmetric.
Inserting \eqref{eq:feedb-cotr-1-spin2-1c}:
\[
\dot V = -k \left( i \bmrho_{\rm d} ^T \ad_{\lambda_1} \bmrho \right)^2  \leqslant 0 .
\]
The system \eqref{eq:contr-liouv-1spin} with the feedback \eqref{eq:feedb-cotr-1-spin2-1c} is time-varying and it is not possible to get rid completely of the time dependence by considering the error system $ {\bm e} = \bmrho_{\rm d} -  \bmrho $:
\[
\dot {\bm e } = -i \ad_{H_{\rm f}} {\bm e } - i u \ad_{H_{\rm c, ns}} ( {\bm e } - \bmrho_{\rm d} ).
\]
However, from \eqref{eq:dotV-1spin-1c}, $ \dot{V} $ is time-indepedent thanks to the cancellations and so must be $ V $ (its time-dependence is only apparent).
Hence we can use LaSalle invariance principle for autonomous systems and the positive limit set of $ \bmrho (t)$ is still the largest invariant set, call it $ {\cal E}$, confined to $ {\cal N} = \{ \bmrho \text{ such that } \dot V =0 \} $ (and corresponding to $ u =0 $).
To compute $ {\cal E} $, following the same idea of the proof of Theorem~2 of \cite{Jurdjevic5}, in $ {\cal N} $ it must also be $ \de{u}{t} =0$.
Explicitely:
\beqa
\de{u}{t} & = & -i k \left( \dot{\bmrho}_{\rm d} ^T \ad_{\lambda_1} \bmrho +  \bmrho_{\rm d} ^T \ad_{\lambda_1} \dot{\bmrho} \right) \label{eq:du-comp-1spin}  \\
& = &  k  h^3 \bmrho_{\rm d} ^T [  -i \ad_{\lambda_1},  \, -i \ad_{\lambda_3} ] \bmrho = \sqrt{2}  k  h^3 i  \bmrho_{\rm d} ^T \ad_{\lambda_2} \bmrho \nonumber.
\eeqa
Notice, however, that $ u = \de{u}{t} =0 $ yield bilinear forms as opposed to the quadratic forms of the original proof of \cite{Jurdjevic5}; that in addition we have the constraint of $ \| \bmrho(t) \| = {\rm const} \neq 0 $ to deal with; and that the Lie algebra involved is composed of only skew-symmetric matrices, which applied to a point yields (out of the singularities) the tangent plane to the sphere $ \mathbb{S}^2_r $ (not $ \mathbb{R}^3 $).
This makes the condition of \cite{Jurdjevic5} nonglobal. 
For example both bilinear forms \eqref{eq:feedb-cotr-1-spin2-1c} and \eqref{eq:du-comp-1spin} are identically zero on the great circles $ \varrho^3_{\rm d} = \varrho^3 =0 $, regardless of the values of $ \varrho^j_{\rm d} $, $ \varrho^j $, $ j=1,2$.
If $ \varrho^3_{\rm d}  \neq 0 $ then in $ {\cal N} $ we have 
\beq
\varrho^j (t) = \frac{\varrho^3}{\varrho^3_{\rm d} } \varrho^j_{\rm d} (t) , \qquad j =1,2 .
\label{eq:rhoj-on-N-1c}
\eeq
and the closed loop system confined to $ {\cal N}  $ is given by the drift alone:
\beq
\dot \bmrho = \sqrt{2} h^3 \begin{bmatrix} 
0 \\ - \varrho^2 \\ \varrho^1 \\ 0 \end{bmatrix}
= \sqrt{2} h^3 \frac{\varrho^3}{\varrho^3_{\rm d} } \begin{bmatrix} 
0 \\ - \varrho^2_{\rm d} \\ \varrho^1_{\rm d} \\ 0 \end{bmatrix} .
\label{eq:cl-loop-on-N-1c}
\eeq
The presence of spurious equilibria in $  {\cal N}  $ is equivalent to the feasibility of \eqref{eq:cl-loop-on-N-1c} with $ \varrho^3 \neq \varrho^3_{\rm d} $.
In $  {\cal N} $, expanding the isospectral constraint $ \| \bmrho_{\rm d} \|^2 = \| \bmrho \|^2 $, one gets: 
\begin{equation}
\begin{split}
(\varrho^{1}_{\rm d}) ^2 +  (\varrho^{2}_{\rm d}) ^2 + (\varrho^{3}_{\rm d}) ^2  & =  (\varrho^{1}) ^2 +  (\varrho^{2}) ^2 + (\varrho^{3}) ^2 \\
= \frac{(\varrho^{3}) ^2}{(\varrho^{3}_{\rm d}) ^2} \left( (\varrho^{1}_{\rm d}) ^2 +  (\varrho^{2}_{\rm d}) ^2 \right) +  (\varrho^{3}) ^2   & =  \frac{(\varrho^{3}) ^2}{(\varrho^{3}_{\rm d}) ^2}  \left((\varrho^{1}_{\rm d}) ^2 +  (\varrho^{2}_{\rm d}) ^2 + (\varrho^{3}_{\rm d}) ^2 \right) \\
\end{split}
\label{eq:equip-1spin-1c}
\end{equation}
The first and last expression of \eqref{eq:equip-1spin-1c} are equal if and only if $  \frac{(\varrho^3) ^2}{(\varrho^3_{\rm d}) ^2} =1 $, i.e., $ \varrho^{3} = \pm \varrho^3_{\rm d} $.
While $ \varrho^{3} = + \varrho^3_{\rm d} $ is already on the equilibrium since \eqref{eq:rhoj-on-N-1c} implies that $ \bmrho = \bmrho_{\rm d} $, the case $ \varrho^{3} = - \varrho^3_{\rm d} $ lead to a singular point of the control law given by $ \bmrho(t) = - \bmrho_{\rm d} (t) $ (again from \eqref{eq:rhoj-on-N-1c}).
Such a singularity corresponds to the antipodal point to the current desired position and is an isolated unstable equilibrium point.
Hence, whenever $ \varrho^3_{\rm d} \neq 0 $, $ {\cal E} = \{ \bmrho(t) = \pm \bmrho_{\rm d} (t) \} $ and the closed loop system almost globally tracks the desired orbit in an asymptotically stable manner.
Since $ V  $ is almost always decreasing along the trajectories of the closed loop system and the condition $ \varrho^3_{\rm d} \neq 0 $ does not depend on time, it is possible to recompute $ {\cal E} $ in terms of the initial conditions:
\beq
\begin{split}
u (t) = &  -i k \bmrho^T_{\rm d} (0) e^{ it h^3 \ad_{\lambda_3} } \ad_{\lambda_1} e^{- it h^3 \ad_{\lambda_3} } \bmrho(0) \\
= &  \sqrt{2} k \bmrho^T_{\rm d} (0) 
\begin{bmatrix} 
0 & 0 & 0 & 0 \\
0 & 0 & 0 & - \sin (\sqrt{2} t h^3) \\
0 & 0 & 0 & - \cos (\sqrt{2} t h^3) \\
0 & \sin (\sqrt{2} t h^3) & \cos (\sqrt{2} t h^3) & 0 
\end{bmatrix} 
 \bmrho(0) \\
= &  \sqrt{2} k \left( \cos(\sqrt{2} t h^3) \left(  \varrho^3_{\rm d}(0) \varrho^2(0)  - \varrho^2_{\rm d}(0)  \varrho^3 (0) \right)  
\right. \\ & \left.  
+ \sin (\sqrt{2} t h^3) \left(  \varrho^3_{\rm d}(0) \varrho^1(0)  - \varrho^1_{\rm d}(0)  \varrho^3 (0) \right) \right) =0
\end{split}
\label{eq:feedb-comp-1spin-1c}
\eeq
leads to conditions equivalent to \eqref{eq:rhoj-on-N-1c} and \eqref{eq:equip-1spin-1c} in terms of $ \bmrho (0) $ and $ \bmrho_{\rm d} (0) $.
Hence $ \bmrho(t) = - \bmrho_{\rm d} (t) $ if and only if $ \bmrho(0) = - \bmrho_{\rm d} (0) $, i.e., the system is initialized in the antipodal point with respect to the initial condition $ \bmrho_{\rm d} (0) $ of the desired trajectory.

\qed

\begin{remark}
The closed loop system is a nonlinear Bloch equation very similar to those obtained by considering radiation damping effects \cite{Abergel1}.
\end{remark}
\begin{remark}
A periodic solution of an autonomous system can never be a global attractor of a compact set (like the sector of the sphere delimited by $ \varrho^3_{\rm d} \neq 0 $ and not containing $ -\bmrho_{\rm d} (0)$).
The time-varying formulation allows to bypass this topological obstruction.
\end{remark}
\begin{remark}
The exact cancellation of the drift in \eqref{eq:dotV-1spin-1c} is crucial for the proof of stability.
If $ h^3_{\rm d} \neq h^3 $, in fact, \eqref{eq:dotV-1spin-1c} is not homogeneous in $ u$ and the set of singular points in $ {\cal N} $ is larger.
\end{remark}

\begin{remark}
If rather than tracking a given orbit (a full state stabilization problem) we are interested only in the orbital asymptotic stabilization of the invariant set $ \varrho^3_{\rm d} = {\rm const}$, then the problem becomes unidimensional.
In this case, simple distance-like feedback laws like
\beq
u  =  k \varrho^2 ( \varrho^3_{\rm d} - \varrho^3 )
\label{eq:feedb-law1-1c}
\eeq
readily provide a solution.
\end{remark}

\begin{figure}[ht]
\begin{center}
 \includegraphics[width=6cm]{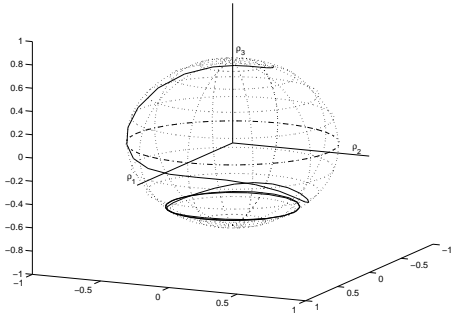}
 \caption{Closed-loop trajectory on the Bloch sphere for the controller \eqref{eq:feedb-cotr-1-spin2-1c}.}
\label{fig:ex1qubit1a}
\end{center}
\end{figure}

\begin{figure}[h]
\begin{center}
 \includegraphics[width=5cm]{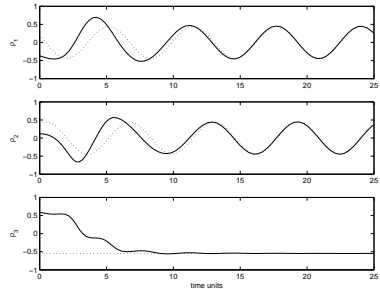}
 \caption{The components of the Bloch vector of $ \bmrho_{\rm d}(t) $ (dotted line) and $ \bmrho(t) $ (solid line) for the same data as in Fig.~\ref{fig:ex1qubit1a}.}
\label{fig:ex1qubit1b}
\end{center}
\end{figure}

\begin{figure}[h]
\begin{center}
 \includegraphics[width=5cm]{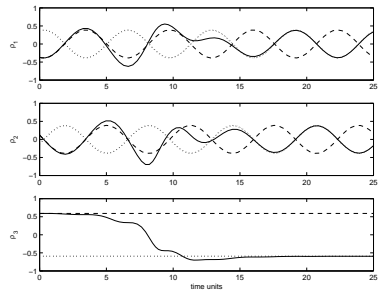}
 \caption{The closed-loop trajectories of the system with intial state (dashed line) ``antipodal'' to the desired state (dotted line) $ \bmrho_{\rm d} (0) = - \bmrho(0) $ and from the same antipodal initial state plus a small perturbation (solid line).}
\label{fig:ex1qubit1c}
\end{center}
\end{figure}

In Figg.~\ref{fig:ex1qubit1a}-\ref{fig:ex1qubit1c}, simulations of the closed loop system with the controller \eqref{eq:feedb-cotr-1-spin2-1c} are shown.
In Fig.~\ref{fig:ex1qubit1c} the instability of the antipodal point is shown: while $ \bmrho(0) = - \bmrho_{\rm d} (0) $ implies the state (dashed line) is not converging to $ \bmrho_{\rm d} (t)$ (dotted line), a small perturbation in $ \bmrho(0) $ is enough to make $ \bmrho(t) $ (solid line) converging to $ \bmrho_{\rm d} (t)$.

When the rf field rotates at the Larmor frequency, the Hamiltonian of the system is driftless as $ h^3 = \omega_o - \omega_{\rm rf} =0$.
In this case, in the rotating frame we have a nutation motion, i.e., just a rotation around the $ \lambda_1 $ axis: 
\beq
\dot \bmrho =-i u \ad_{\lambda_1} \bmrho .
\label{eq:nutation-1-spin}
\eeq
Assuming $ \bmrho_{\rm d} (t) = \bmrho_{\rm d} (0)$, then both the feedback laws \eqref{eq:feedb-law1-1c} and \eqref{eq:feedb-cotr-1-spin2-1c} still hold and give a time-independent closed loop system.
As a consequence, almost global stabilization is no longer achievable on a compact and in fact both schemes correspond to partial state stabilization schemes, since in \eqref{eq:nutation-1-spin} $ \dot \varrho^1 =0 $ remains critically stable in closed loop.

%%%%%%%%%%%%%%%%%%%%%%%%%%%%%%%%%%%%%%%%%%%%%%%%%%%%%%%%%%%%%%%%%%%%%%%%%%%%%%

\subsection{State feedback for 2 spin-$ \frac{1}{2} $ systems}
\label{sec:2spin}
Assuming the entire two spin $ 1/2 $ state tensor $ \bmrho $ is available on-line, we want now to generalize the feedback law of Section~\ref{sec:one-spin-feedb-laws} to the ensembles of weakly coupled spin $ 1/2 $ described above.
Only the nonselective case will be treated.
Clearly, the selective Hamiltonian of eq.~\eqref{eq:Ham-2spin-c} has more control authority available, hence its stabilization is an easier task than the nonselective case.
As before, let $ \bmrho_\alpha$, $ \bmrho_\beta $ and $  \bmrho_{\alpha_{\rm d}}$, $ \bmrho_{\beta_{\rm d}} $ be the reduced densities respectively of $ \bmrho $ and $ \bmrho_{\rm d} $.
An obvious prerequisite for stabilizability is that $ \bmrho $ and $ \bmrho_{\rm d} $ both belong to $ {\cal S} $, which  implies for example that $ \| \bmrho\| = \| \bmrho_{\rm d} \| $.
We shall also assume that the intial condition is a product state $ \bmrho (0) = \bmrho_\alpha (0)\ox \bmrho_\beta(0) $.

The proof of Proposition~\ref{prop:orb-tr-1spin-1c} was essentially relying on the Jurdjevic-Quinn sufficient condition for stabilizability \cite{Jurdjevic5}, opportunely modified in order to deal with skew-symmetric infinitesimal generators.
The key tool is the so-called ``ad-condition'', i.e., a particular type of Lie brackets often used for testing controllability, see \cite{Cla-contr-root1}.
The following proposition shows that this condition is never satisfied by the system \eqref{eq:liov-2spin}.

\begin{lemma}
\label{lemma:Jurd-Qu-not}
For the system \eqref{eq:liov-2spin} 
\beq
{\rm span } \left\{  -i \ad_{H_{\rm f}} , \, -i \ad_{H_{\rm c,ns}} , \, [  -i \ad_{H_{\rm f}} , \, -i \ad_{H_{\rm c,ns}}] ,\ldots,  [  -i \ad_{H_{\rm f}} ,  \ldots ,[ -i \ad_{H_{\rm f}} ,\,  -i \ad_{H_{\rm c,ns}}]\ldots ] \right\} \neq  \ad_{\mathfrak{g}_{2s}}.
\label{eq:Jurd-Quinn-ad-NOT}
\eeq 
\end{lemma}

\proof
Given any $ A, B \in\mathfrak{g}_{2s} $, for the adjoint representation 
\beq
[ \ad_A, \, \ad_B ] = \ad_{[ A, \, B ]} ,
\label{eq:ad-comm-isom1}
\eeq
with $ \ad_A, \ad_B \in \ad_{\mathfrak{g}_{2s}}$.
Hence, \eqref{eq:Jurd-Quinn-ad-NOT} holds if and only if
\beq
{\rm span } \left\{ -i {H_{\rm f}} , \, -i {H_{\rm c,ns}} , \, [  -i {H_{\rm f}} , \, -i {H_{\rm c,ns}}] ,\ldots,  [  -i {H_{\rm f}} ,  \ldots ,[ -i {H_{\rm f}} ,\,  -i {H_{\rm c,ns}}]\ldots ] \right\} \neq  {\mathfrak{g}_{2s}}.
\label{eq:Jurd-Quinn-NOT}
\eeq 
Computing explicitly the first Lie brackets:
\begin{subequations}
\label{eq:lie-br-2spin}
\beqa
&&  [  -i {H_{\rm f}} , \, -i {H_{\rm c, ns}}]  =  -i \left( 
   h^{03} \Lambda_{02} 
-  h^{33} \Lambda_{32} 
+  h^{30} \Lambda_{20} 
-  h^{33} \Lambda_{23} \right)  \label{eq:lie-br-2spin-a} \\
&& [  -i {H_{\rm f}} ,  [  -i {H_{\rm f}} , \, -i H_{\rm c, ns}]]  =   -i \left( 
  \left( ( h^{33})^2 -  ( h^{03})^2 \right)  \Lambda_{01} 
+ \left( ( h^{33})^2 -  ( h^{30})^2 \right) \Lambda_{10} \right) 
\label{eq:lie-br-2spin-b}
\eeqa
\end{subequations}
Comparing \eqref{eq:Ham-2spin-b} and \eqref{eq:lie-br-2spin-b} it is clear what the pattern of iterated Lie brackets will be, since the basis elements involved are the same.
In particular, the basis directions $ -i \Lambda_{11} $, $ -i \Lambda_{12} $, $ -i \Lambda_{21} $ and $ -i \Lambda_{22} $ are never touched by such sequences of commutators.
Hence the claim \eqref{eq:Jurd-Quinn-NOT} or, equivalently, \eqref{eq:Jurd-Quinn-ad-NOT}.
\qed

As for the single spin $ 1/2 $ case, we want to achieve asymptotically stable tracking of the following periodic orbit:
\beq
\begin{split}
\dot{\bmrho}_{\rm d}  = & -i \ad_{H_{\rm f_{\rm d}} } \bmrho_{\rm d} \\
= & -i \left( h^{03}_{\rm d} \ad_{\Lambda_{03}} + h^{30}_{\rm d} \ad_{\Lambda_{30}} + h^{33}_{\rm d} \ad_{\Lambda_{33}} \right) \bmrho_{\rm d} , \qquad \bmrho_{\rm d}(0) \in {\cal S}
\end{split}
\label{eq:ref-orb-2spin}
\eeq
by means of a single control input.

\begin{proposition}
\label{prop:orb-tr-2spin-1c}
Whenever $ H_{\rm f} $ is $ H_{\rm c,ns} $-strongly regular, the feedback 
\beq
u =  k  \llangle\bmrho_{\rm d} , \, -i \ad_{H_{\rm c, ns}}  \bmrho \rrangle  
\label{eq:feedb-orb-tr-2spin-1c}
\eeq
with $ k \in \mathbb{R}^+ $, asymptotically stabilizes the system \eqref{eq:liov-2spin} to the time-varying reference state $ \bmrho_{\rm d} (t) $ given by \eqref{eq:ref-orb-2spin} with $H_{\rm f_{\rm d}} = H_{\rm f} $, for all $ \bmrho(0) \in {\cal S}$  with the exception of $ \bmrho(0) $ such that  $ (\bmrho_\alpha (0) , \,  \bmrho_\beta(0) ) = - (\bmrho_{\alpha_{\rm d}} (0) , \,  \bmrho_{\beta_{\rm d}} (0) ) $ and of all pairs $ ( \bmrho(0), \, \bmrho_{\rm d} (0) ) $ having  $ (\varrho^3_\alpha , \,  \varrho^3_{\alpha_{\rm d}} ) = (0,\, 0 )$ and $ (\varrho^3_\beta , \,  \varrho^3_{\beta_{\rm d}} ) = (0, \, 0 )  $.
\end{proposition}

\proof
As in the proof of Proposition~\ref{prop:orb-tr-1spin-1c}, take as Lyapunov function the analogous of the distance \eqref{eq:dist-1spin}, $ V (t) = \| \bmrho_{\rm d} \|^2 - \llangle \bmrho_{\rm d} (t), \, \bmrho (t) \rrangle $.
Again, when differentiating the drift disappears,
\begin{subequations}
\label{eq:dot-V-orb-tr}
\beqa
\dot{V}  & = & - \llangle \dot{\bmrho}_{\rm d} , \, \bmrho  \rrangle -  \llangle \bmrho_{\rm d} , \, \dot{\bmrho}  \rrangle \label{eq:dot-V-orb-tr-1} \\
& = & i \left( \llangle \ad_{H_{\rm f}} \bmrho_{\rm d} , \,\bmrho  \rrangle + \llangle \bmrho_{\rm d} , \,( \ad_{H_{\rm f}} + u \ad_{H_{\rm c, ns}} )  \bmrho  \rrangle \right) \label{eq:dot-V-orb-tr-2} \\
& = & -  u  \llangle \bmrho_{\rm d} , \,-i \ad_{H_{\rm c, ns}} \bmrho  \rrangle ,\label{eq:dot-V-orb-tr-3}
\eeqa
\end{subequations}
and $ \dot{V} $ is made negative semidefinite by the choice of feedback \eqref{eq:feedb-orb-tr-2spin-1c}.
Complications arise when checking the emptiness of the invariant set $ {\cal E} $ in $ {\cal N}  = \{ \bmrho \text{ s.t. } \dot{V} =0 \} $.
In fact, from Lemma~\ref{lemma:Jurd-Qu-not}, the Jurdjevic-Quinn condition never applies to tensor product systems.
Furthermore, since the reduced densities $ \bmrho_\alpha (t) $ and $ \bmrho_\beta (t) $ in $ {\cal N}$ have time-varying norm, neither the method used in the proof of Proposition~\ref{prop:orb-tr-1spin-1c} is directly applicable.
However, it can be applied ``stroboscopically'' i.e., at the time instants $ t= c \tau_p $, $ c \in \mathbb{N}$, in which the nonlocal part of the dynamics disappears.
From \eqref{eq:feedb-orb-tr-2spin-1c}:
\beqan
u & = & -i k  \bmrho_{\rm d}^T  \ad_{H_{\rm c, ns}}  \bmrho  \\
\de{u}{t} & = & -i k \left( \dot{\bmrho} ^T_{\rm d} \ad_{H_{\rm c, ns}}  \bmrho  +\bmrho ^T_{\rm d} \ad_{H_{\rm c, ns}}  \dot{\bmrho } \right) \\
 & = &  -i k \bmrho ^T_{\rm d} [ \ad_{H_{\rm c, ns}} , \, \ad_{H_{\rm f}} ]  \bmrho  \\
 & = &  -i k \bmrho ^T_{\rm d} \ad_{ [ H_{\rm c, ns} , \, H_{\rm f} ] } \bmrho 
\eeqan
where we have used \eqref{eq:ad-comm-isom1}.
The explicit expression for $ \ad_{ [ H_{\rm c, ns} , \, H_{\rm f} ] } $ follows applying \eqref{eq:ad-comm-isom1} to \eqref{eq:lie-br-2spin-a}.
From \eqref{eq:comm-drift-2spin} and \eqref{eq:int-exp-drift-2spin}, in ${\cal N} $ one can write
\beq
\begin{split}
u (t) = &  -i k \left(  \bmrho^T_{\rm d} (0) e^{ it h^{33} \ad_{\Lambda_{33}} } e^{ it h^{03} \ad_{\Lambda_{03}} } \ad_{\Lambda_{01}}  e^{-it h^{03} \ad_{\Lambda_{03}} } e^{-it h^{33} \ad_{\Lambda_{33}} } \bmrho(0)
\right. \\ & \left. 
+  \bmrho^T_{\rm d} (0) e^{ it h^{33} \ad_{\Lambda_{33}} } e^{ it h^{30} \ad_{\Lambda_{30}} } \ad_{\Lambda_{10}}  e^{-it h^{30} \ad_{\Lambda_{30}} } e^{-it h^{33} \ad_{\Lambda_{33}} } \bmrho(0)  \right) =0 ,
\end{split}
\label{eq:feedb-2spin-1c}
\eeq
and similarly for $ \de{u}{t} =0 $.
At $ t= c \tau_p $, $ e^{ \pm it h^{33} \ad_{\Lambda_{33}} } = I_4 $, hence \eqref{eq:feedb-2spin-1c} contains only local terms and the integration in $ \bmrho $, $ \bmrho_{\rm d} $ resembles \eqref{eq:feedb-comp-1spin-1c}:
\beq
\begin{split}
\left. u (t) \right|_{t=c \tau_p}  = & -i\sqrt{2}  k \left(  
  \bmrho^T_{\rm d} (0) I_4 \ox e^{ it h^{03} \ad_{\lambda_3} } \ad_{\lambda_1} e^{- it h^{03} \ad_{\lambda_3} } \bmrho(0) 
+  \bmrho^T_{\rm d} (0)e^{ it h^{30} \ad_{\lambda_3} } \ad_{\lambda_1} e^{- it h^{30} \ad_{\lambda_3} } \ox I_4  \bmrho(0) \right) \\
= & 2  k \left( \bmrho_{d_\alpha}^T (0)  \bmrho_\alpha (0) \ox 
\bmrho^T_{d_\beta} (0) 
\begin{bmatrix} 
0 & 0 & 0 & 0 \\
0 & 0 & 0 & - \sin ( t h^{03}) \\
0 & 0 & 0 & - \cos ( t h^{03}) \\
0 & \sin ( t h^{03}) & \cos ( t h^{03}) & 0 
\end{bmatrix} 
 \bmrho_\beta (0)
\right. \\ & \left. 
+ \bmrho^T_{d_\alpha} (0) 
\begin{bmatrix} 
0 & 0 & 0 & 0 \\
0 & 0 & 0 & - \sin ( t h^{30}) \\
0 & 0 & 0 & - \cos ( t h^{30}) \\
0 & \sin ( t h^{30}) & \cos ( t h^{30}) & 0 
\end{bmatrix} 
 \bmrho_\alpha (0) \ox  \bmrho_{d_\beta}^T (0)  \bmrho_\beta (0) \right) =0 .
\end{split}
\label{eq:feedb-comp-2spin-1c}
\eeq
Eq. \eqref{eq:feedb-comp-2spin-1c} is enough to show that the structure of $ {\cal E} $ at $ t= c \tau_p $ reflects essentially that of Proposition~\ref{prop:orb-tr-1spin-1c}.
In fact, the assumption of $ H_{\rm c, ns} $-strong regularity of $ H_{\rm f}$ implies that (excluding the case $ \bmrho_{\alpha_{\rm d}} ^T (0) \bmrho_\alpha (0) = \bmrho_{\beta_{\rm d}} ^T (0) \bmrho_\beta (0) = 0 $ corresponding to the antipodal case, but only for pure reduced densities), each of the two terms of \eqref{eq:feedb-comp-2spin-1c} must be zero.
Therefore in $ {\cal N} $ we have the following invariant states (for $ t= c \tau_p $):
\begin{enumerate}
\item $ \bmrho_\alpha (0) = - \bmrho_{\alpha_{\rm d}} (0) $ and to $ \bmrho_\beta (0) = - \bmrho_{\beta_{\rm d}} (0) $ (``local'' antipodal case); 
\item $ \varrho^3_\alpha = \varrho^3_{\alpha_{\rm d}} =0 $ and/or $ \varrho^3_\beta = \varrho^3_{\beta_{\rm d}} =0 $, regardless of the values of the other terms $ \varrho^{0j}$, $ \varrho^{j0}$, $ \varrho^{0j}_{\rm d}$, $ \varrho^{j0}_{\rm d}$, $ j=1,2$ (reduced densities of true and desired trajectories are horizontal great circles, for any degree of mixing, including the completely mixed case).
\end{enumerate}
All cases above correspond to $ u =0$ and give the invariant set $ {\cal E} $ at $ t= c \tau_p $. 
Hence, for any $t$, $ {\cal E} $ will contain at most a subset of them.
All are excluded by the assumptions of the Proposition.
\qed

% Alternatively, Proposition~\ref{prop:orb-tr-2spin-1c} can be proved using the foliation structure determined by \eqref{eq:int-Isin-2spin} and \eqref{eq:inv-man-drift}.
% Since this has to be preserved by the trajectories in $ {\cal N}$, a reduced version of the Jurdjevic-Quinn condition is applicable, in which the $ \ad$-brackets are required to generate only the Lie subalgebras acting on the 3-dimensional spheres determined by \eqref{eq:inv-man-drift}.

\begin{figure}[h]
\begin{center}
\includegraphics[width=7cm]{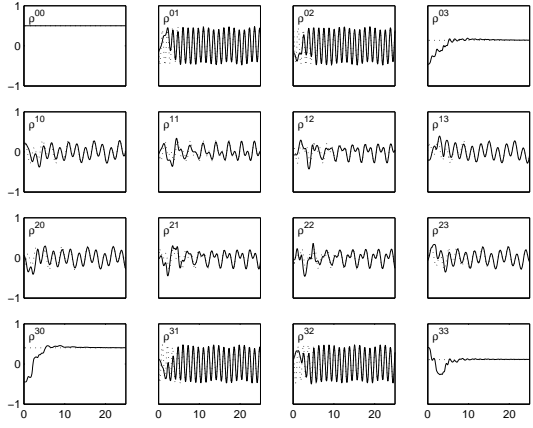}
 \caption{The 16-components of the tensor $ \bmrho_{\rm d} $ (dotted lines) and $ \bmrho$ (solid line) for the orbit tracking problem.}
\label{fig:16-tensor-orb-track-2spin}
\end{center}
\end{figure}

\begin{figure}[h]
\begin{center}
\includegraphics[width=5cm]{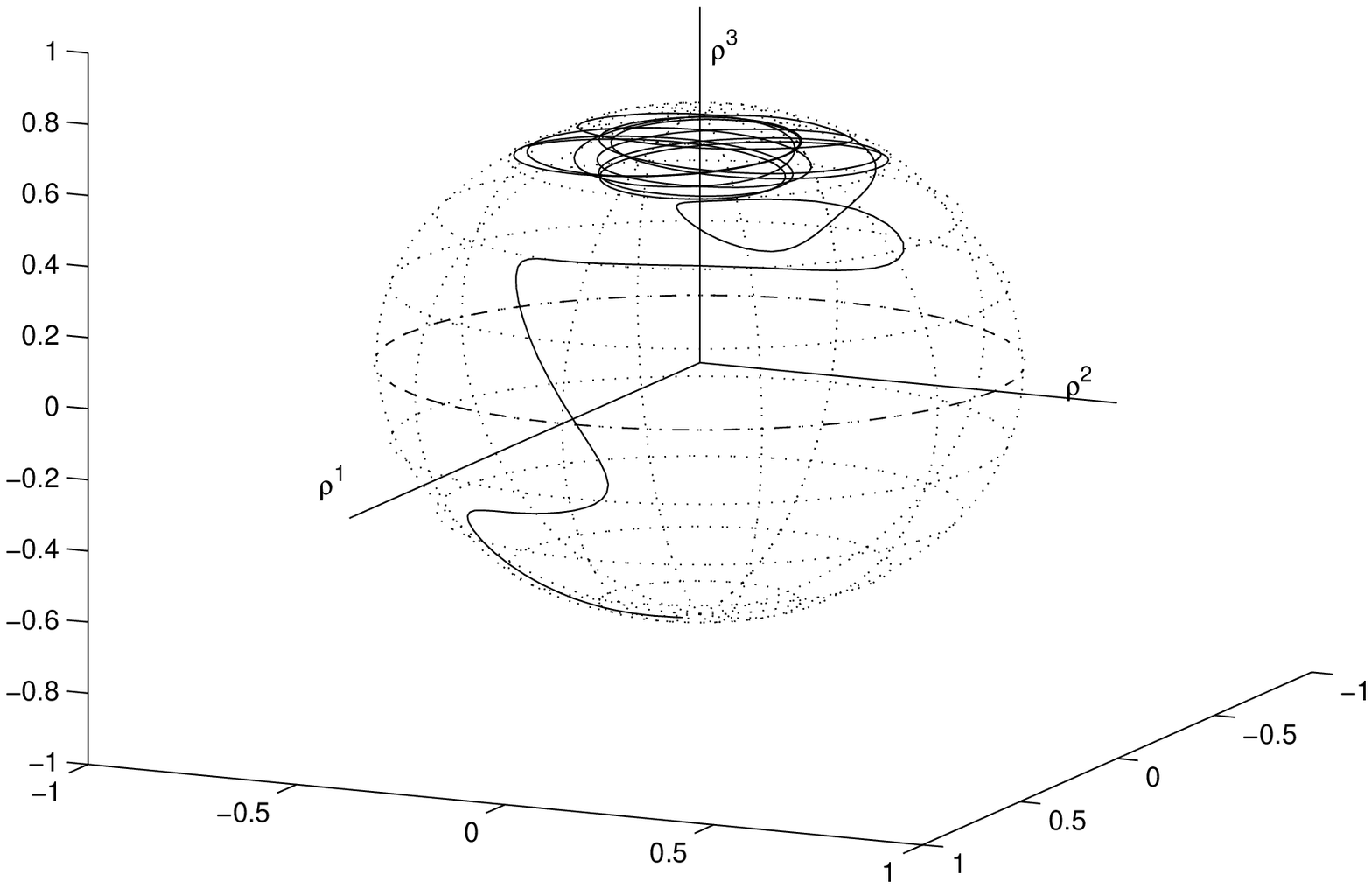}
\includegraphics[width=5cm]{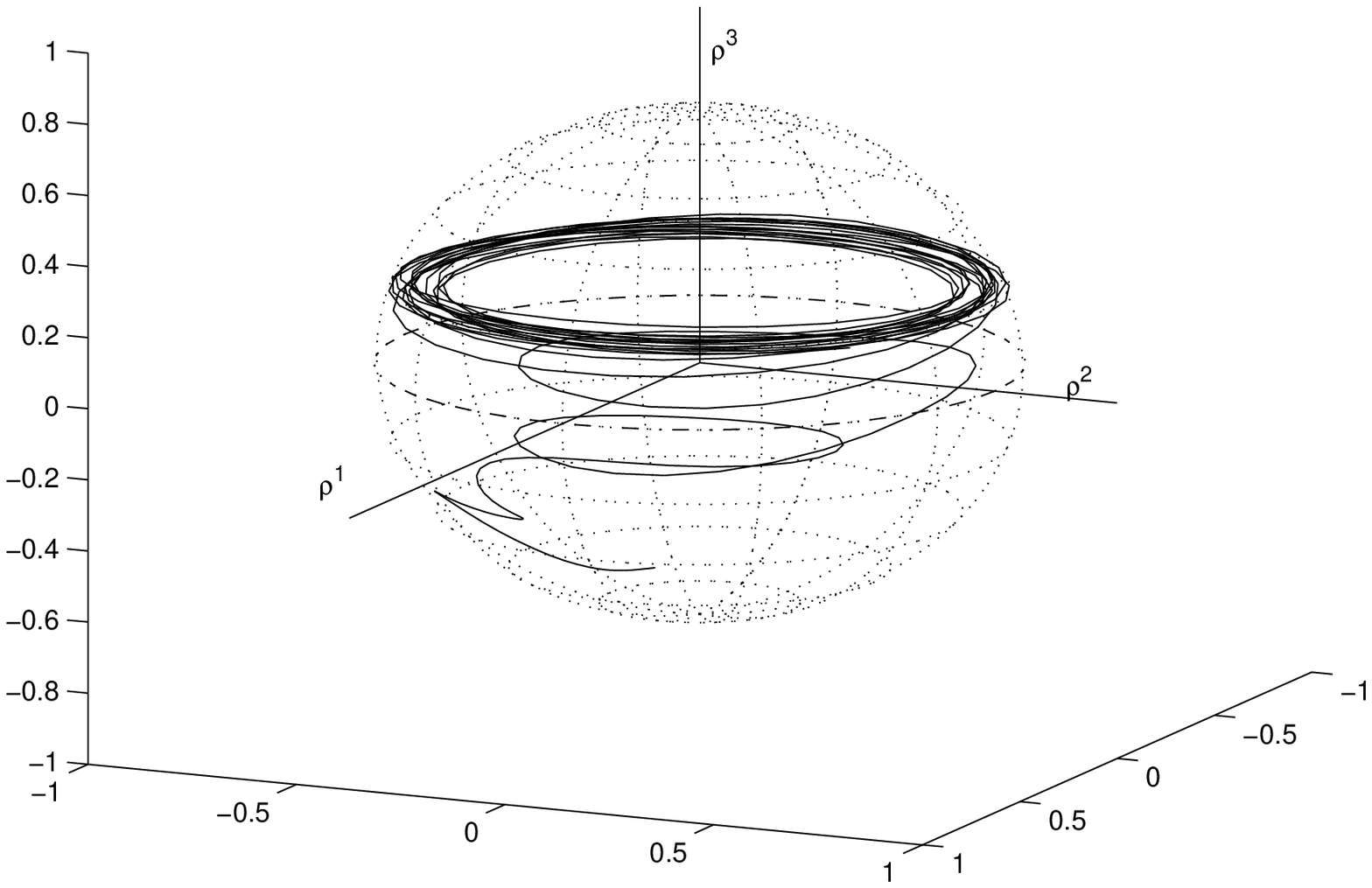}
 \caption{Closed-loop reduced densities $ \bmrho_\alpha (t) $ (left) and $ \bmrho_\beta (t) $ (right) corresponding to Fig.~\ref{fig:16-tensor-orb-track-2spin}.}
\label{fig:orb-track-red-2spin}
\end{center}
\end{figure}

For a typical simulation, the entire 16-state reference tensor (dotted) and the tensor of the closed-loop system (solid) are shown in Fig.~\ref{fig:16-tensor-orb-track-2spin}, and the two reduced densities in  Fig.~\ref{fig:orb-track-red-2spin}.

\begin{remark}
\label{rem:local-feedb-2spin} 
Since $ -i \ad_{H_{\rm c,ns} } $ is local, the feedback action \eqref{eq:feedb-orb-tr-2spin-1c} preserves $ \| \bmrho_\alpha \| $ and $ \| \bmrho_\beta \|$.
\end{remark}

\begin{remark}
\label{rem:2control-2spin}
From a practical point of view, although out of the singular set convergence is guaranteed, the rate of convergence may be very slow for inital conditions near the singularities.
Having a second nonselective control field at the same frequency $ \omega_{\rm rf} $ but with a phase difference of $ 90^\circ $ i.e., using 
\[
H_{\rm c } = u_1 ( \Lambda_{01} + \Lambda_{10} ) +  u_2 ( \Lambda_{02} + \Lambda_{20} )
\]
in place of \eqref{eq:Ham-2spin-b}, with the feedback law $ u_2 = \llangle \bmrho_{\rm d} , \, -i \ad _{(\Lambda_{02} + \Lambda_{20} )} \bmrho \rrangle $ helps considerably in improving the convergence rate.
\end{remark}

In the selective case described by the Hamiltonian \eqref{eq:Ham-2spin-a}  and \eqref{eq:Ham-2spin-c}, the two feedback laws for $ u_{01} $ and $ u_{10} $ are of the same type as \eqref{eq:feedb-orb-tr-2spin-1c}.
Tuning the two rf fields exactly on the two Larmor frequencies $ \omega_{o,\alpha}$ and $ \omega_{o,\beta}$, one gets $ h^{03} = h^{30} =0$.
Therefore the proof of Proposition~\ref{prop:orb-tr-2spin-1c} does not apply straightforwardly.
However, the idea behind the proof still holds true, namely that the two reduced closed-loop evolutions are nonidentical.
The presence of two controls in the selective Hamiltonian \eqref{eq:Ham-2spin-c} emphasizes this distinguishability and simplifies the stabilizability argument considerably with respect to Proposition~\ref{prop:orb-tr-2spin-1c}.

Another consequence of this ``distinguishability'' notion, as a generalization of the argument in the proof of Proposition~\ref{prop:orb-tr-2spin-1c}, is that the same tracking scheme can be used for more complicated free Hamiltonians than \eqref{eq:Ham-2spin-a}.
In particular, any 
\beq
H_{\rm f} = h^{03} \Lambda_{03} +  h^{30} \Lambda_{30} + h^{jk} \Lambda_{jk} , \qquad j, \, k \neq 0 
\label{eq:free-Ham-complic-2spin}
\eeq
can be used in Proposition~\ref{prop:orb-tr-2spin-1c}, provided that $ H_{\rm f_{\rm d}} = H_{\rm f}$.
We state it as a Corollary.

\begin{corollary}
\label{cor:orb-track-complic-2spin}
If $ H_{\rm f} $ given by \eqref{eq:free-Ham-complic-2spin} is $ H_{\rm c,ns} $-strongly regular, the feedback law \eqref{eq:feedb-orb-tr-2spin-1c} asymptotically stabilizes the system $ \dot \bmrho = -i \ad_{(H_{\rm f} + u H_{\rm c, ns})} \bmrho $ to the time-varying reference state $ \bmrho_{\rm d} (t) $ given by $ \dot \bmrho_{\rm d} = -i \ad_{H_{\rm f} } \bmrho_{\rm d} $ for all $ \bmrho(0) \in {\cal S} $, except for the same singular set described in Proposition~\ref{prop:orb-tr-2spin-1c}.
\end{corollary}

%%%%%%%%%%%%%%%%%%%%%%%%%%%%%%%%%%%%%%%%%%%%%%%%%%%%%%%%%%%%%%%%%%%%%%%%%%

\subsection{$n$ spin $ 1/2 $ case} 

For the Ising Hamiltonian, if we start from a product state, label the spins as $ \alpha $, $\ldots $, $ \nu $ and call $ \rho_\alpha $ , $ \ldots $, $ \rho_\nu $ the corresponding reduced densities, then we can arrive at the same conclusion as Proposition ~\ref{prop:orb-tr-2spin-1c}, provided all energy levels are neither equal nor equispaced.
The proof is omitted as it makes use of the same techniques used above, only the notation is more cumbersome.

\begin{proposition}
If $ H_{\rm f} $ is $ H_{\rm c,ns} $-strongly regular, the feedback
\beq
u = k \llangle  \bmrho_{\rm d}, \, -i   \ad_{H_{\rm c,ns} } \bmrho \rrangle 
\label{eq:feedb-n-spin}
\eeq
with $ k \in \mathbb{R}^+ $, asymptotically stabilizes the system $ \dot \bmrho = -i ( \ad_{H_{\rm f}} + u \ad_{H_{\rm c,ns} }) \bmrho$ to the time-varying reference orbit $ \bmrho_{\rm d} (t) $ given by $ \dot \bmrho_{\rm d} = -i  \ad_{H_{\rm f}} \bmrho_{\rm d}  $  , for all $ \bmrho(0)$ such that $ \| \bmrho (0) \| = \| \bmrho_{\rm d} (0) \|$, with the exception of the antipodal point $ (\bmrho_\alpha (0) , \ldots ,  \bmrho_\nu (0) ) = - (\bmrho_{\alpha _{\rm d}} (0) , \ldots ,  \bmrho_{\nu_{\rm d}} (0) ) $ and of all pairs $ ( \bmrho(0), \, \bmrho_{\rm d} (0) ) $ having  $ (\varrho^3_\alpha , \,  \varrho^3_{\alpha_{\rm d}} ) = (0,\, 0 )$, $\ldots $, $ (\varrho^3_\nu , \,  \varrho^3_{\nu_{\rm d}} ) = (0, \, 0 )  $.
\end{proposition}

%%%%%%%%%%%%%%%%%%%%%%%%%%%%%%%%%%%%%%%%%%%%%%%%%%%%%%%%%%%%%%%%%%%%%%%%%%%%%%%

\section{Suppressing unwanted weak couplings}
\label{sec:suppr}
The $J$-coupling used in the previous Sections is an indirect coupling mechanics, physically due to the electrons shared in the chemical bonds between the atoms.
Apart from this coupling, there are other interaction mechanisms, due to the direct or electron-mediated interactions between the spins.

Consider the nonselective two spin $ 1/2 $ Hamiltonian \eqref{eq:free-Ham-complic-2spin} and 
\beq
H_{\rm f_{\rm d}} = h^{03} \Lambda_{03} +  h^{30} \Lambda_{30} +  h^{jk}_{\rm d}  \Lambda_{jk} , \qquad j,k \neq 0 .
\label{eq:Hfd-multicoupl-2spin}
\eeq
Call $H_\delta = H_{\rm f_{\rm d}} - H_{\rm f} $ the difference between the desired and the true Hamiltonian.
We are interested in treating the extra terms $ H_\delta $ as disturbances and suppressing them by means of feedback.

Assume the frequencies $ h^{jk}_{\rm d} $ are of the same order of magnitude, call it $ 2 \pi /\tau_{\rm d}$.
When the frequency of the disturbance $ H_\delta $, call it $ 2 \pi /\tau_\delta$ ($\simeq h^{jk} _\delta $), is much smaller than $  2 \pi /\tau_{\rm d}$, then $ H_\delta $ can be suppressed by the control action.

\begin{proposition}
\label{prop:dist-rej-2spin}
Assume $ H_{\rm f} $ is $ H_{\rm c,ns} $-strongly regular and $ \tau_\delta >>  \tau_{\rm d}$.
Then there exists a $ \omega_{\rm rf} $ and a sufficiently high gain $ k $ such that the system $ \dot \bmrho = -i \ad_{(H_{\rm f} + u H_{\rm c,ns})} \bmrho $ with the feedback \eqref{eq:feedb-orb-tr-2spin-1c} can track the reference trajectory $ \bmrho_{\rm d} (t) $ given by $ \dot \bmrho_{\rm d} = -i \ad_{H_{\rm f_{\rm d}} } \bmrho_{\rm d} $ and reject the disturbance $ H_\delta $ up to a bounded error.
% of the order $ \tau_{\rm d} / \tau_\delta $.
\end{proposition}

\proof
Since $ H_{\rm f_{\rm d}} \neq H_{\rm f} $, in the proof of Proposition~\ref{prop:orb-tr-2spin-1c} the derivative of the Lyapunov function is no longer homogeneous in  the control. Inserting the feedback \eqref{eq:feedb-orb-tr-2spin-1c},
\beq
\dot V =  \llangle \bmrho_{\rm d} , \,- i \ad_{H_\delta} \bmrho \rrangle -k 
\llangle \bmrho_{\rm d} , \, - i \ad_{H_{\rm c,ns} } \bmrho \rrangle^2 ,
\label{eq:dot-V-multic-2spin}
\eeq
the first term is in general sign indefinite, implying negative semidefiniteness of $ \dot V $ is not guaranteed.
Consider the set $ {\cal N} = \{ \bmrho \text{ s. t. } \dot{V} =0 \}$.
Since $ k \llangle \bmrho_{\rm d} , \, - i \ad_{H_{\rm c,ns} } \bmrho \rrangle^2 \geqslant 0 $, for an invariant set to belong to $ {\cal N}$ it must be $ \llangle \bmrho_{\rm d} , \,- i \ad_{H_\delta} \bmrho \rrangle \geqslant 0 $.
As $ -i \ad_{H_\delta} $ is skew-symmetric, the only solution is  $ \llangle \bmrho_{\rm d} , \,- i \ad_{H_\delta} \bmrho \rrangle = \llangle \bmrho_{\rm d} , \, - i \ad_{H_{\rm c,ns} } \bmrho \rrangle^2 =0$.
This happens for example for the perfect tracking $ \bmrho= \bmrho_{\rm d} $. 
When this is not verified, the steady state is replaced by a limit cycle which is in general stable but not asymptotically stable.

The first term in \eqref{eq:dot-V-multic-2spin} is slow, of small amplitude, of period $ \tau_\delta $ (when $u=0$) and of zero average.
Provided $k$ is sufficiently high, the second term has a fast dynamics with respect to the first one and a large amplitude.
Hence, in the time scale $ \tau_{\rm d} $, $ \llangle \bmrho_{\rm d} , \,- i \ad_{H_\delta} \bmrho \rrangle $ can be thought of as frozen, with the controlled term still acting as a damping.
The cancellation of $ H_\delta $ cannot be perfect because there are regions of the state space where the controllable term of \eqref{eq:dot-V-multic-2spin} vanishes, namely where any of the 4 reduced densities $ \bmrho_{\alpha_{\rm d}} $, $ \bmrho_{\beta_{\rm d}} $, $ \bmrho_{\alpha} $, and $ \bmrho_{\beta} $ approaches the $ \lambda_1 $ axis.
Recall (Remark~\ref{rem:local-feedb-2spin}) that the amplitude of the feedback action only depends from the reduced densities.
In absence of a local precession motion, the system leaves this non-convergence region only due to the coupling terms and it is not possible to  guarantee a recovery from the disturbance-induced instability for all $ H_{\rm f_{\rm d}} $. 
However, since $ h^{03} = - ( \omega_{o, \alpha} - \omega_{\rm rf} )$ and $ h^{30} = - ( \omega_{o, \beta} - \omega_{\rm rf} )$, it is always possible to choose $ \omega_{\rm rf} $ so that $ 1/\tau_\delta $ is small compared to $ h^{03} $ and $ h^{30} $.
The effect of the local precessions is to steer the corresponding reduced dynamics, both the desired and the real ones, out of the uncontrollable alignment with the $ \lambda_1 $ axis.
Therefore, since both local closed-loop dynamics evolve fast compared to $ H_\delta $ and so does $ H_{\rm f_{\rm d}} $, the displacement due to the drift term in \eqref{eq:dot-V-multic-2spin} can be rejected in the fast time scale.
Since the disturbance $ H_\delta $ is persistently exciting the system, no steady state is ever reached, but the error remains bounded.
\qed

\begin{figure}[h]
\begin{center}
\includegraphics[width=7cm]{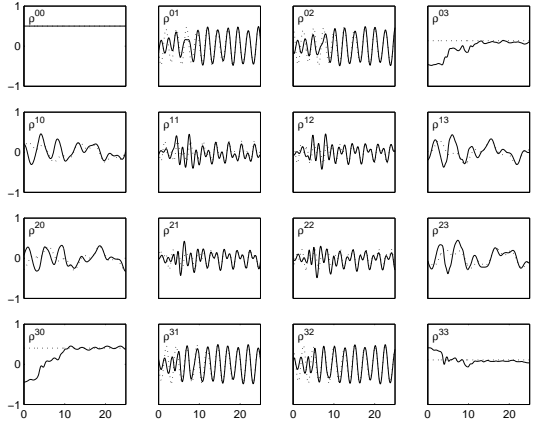}
 \caption{The 16-components of the tensor $ \bmrho_{\rm d} $ (dotted lines) and $ \bmrho$ (solid line) for the weak coupling suppression problem.}
\label{fig:16-tensor-dit-rej-2spin}
\end{center}
\end{figure}

\begin{figure}[h]
\begin{center}
\includegraphics[width=6cm]{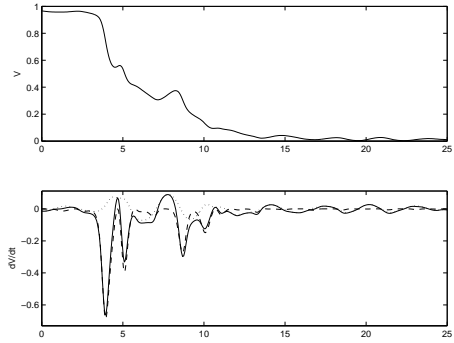}
 \caption{Top: the Lyapunov function $ V $. Bottom: $ \dot{V}$ (solid line) and its 2 components (dotted and dashed lines), see \eqref{eq:dot-V-multic-2spin}.}
\label{fig:16-tensor-dit-rej-2spin-Lyap}
\end{center}
\end{figure}

Two common coupling models often used in the literature are the Heisenberg interaction and the dipole-dipole interaction \cite{Abragam1,Ernst1}.
For example, the dipole-dipole Hamiltonian is given by
\beq
H_{\rm dd} = -\omega_{\rm dd} ( \Lambda_{11} + \Lambda_{22} -2 \Lambda_{33} )
\label{eq:Ham-he-2spin}
\eeq
and models the direct coupling between the magnetic moments in solid state NMR.
As an application of Proposition~\ref{prop:dist-rej-2spin}, we want to modify the response of $ H_{\rm dd} $ so as to reproduce that of a $ J$ coupling.
In the rotating frame $ \omega_{\rm rf} $, assume that the free Hamiltonian is
\beq
H_{\rm f} = h^{03} \Lambda_{03} +  h^{30} \Lambda_{30} +  h^{33} \Lambda_{33} +  h^{11} \Lambda_{11} +  h^{22} \Lambda_{22} 
\label{eq:Ham-dipol-coup-2spin}
\eeq
where $ H_\delta= h^{11} \Lambda_{11} + h^{22} \Lambda_{22} $, $ h^{11} = h^{22} = - h^{33}/2 = - \omega_{\rm dd} $ and $ 10 \omega_{\rm dd} \simeq  h^{03}  \simeq h^{30} $.
The control Hamiltonian is still nonselective and given by \eqref{eq:Ham-2spin-b}.
A typical closed loop behavior for this choice is shown in Fig.~\ref{fig:16-tensor-dit-rej-2spin}. Even after the offset due to the initial condition is recovered, the system does not quicky reach an unperturbed steady state due to the periodic excitation given be $ H_\delta$.
The Lyapunov function $ V $ and its derivative $ \dot {V}$ are shown in Fig.~\ref{fig:16-tensor-dit-rej-2spin-Lyap}. 
Notice further that Corollary~\ref{cor:orb-track-complic-2spin} implies that the coupling included in $ H_{\rm f_{\rm d}} $ needs not be restricted to the vertical $ \Lambda_{33} $ direction. 
Only the weakness of the unwanted coupling $ H_{\rm f} - H_{\rm f_{\rm d}} $ with respect to $ H_{\rm f_{\rm d}} $ and with respect to the local precession frequencies matters for the disturbance rejection.

\section{Feedback from measurable quantities}
A typical NMR measurement apparatus can provide a collective magnetization measurement in the $ (\lambda_1,\, \lambda_2 ) $ plane.
We assume this measure process can cohexist with the excitation of the coil (actuator signal).
While for a single spin ensemble this, together with $ \| \bmrho \| = {\rm const} $, allows to easily recover the entire state vector, for two spin systems what is measurable depends on the nuclear species we are considering.
In the homonuclear case, the measurement corresponds to the output vector $ y = [y_1 \, y_2 ]^T $, where $y_j (t) = \varrho^{0j} (t) + \varrho^{j0}(t)$.
Since also the controls are nonselective, a feedback from $ y $ cannot stabilize the system.
However, if we can, by means of a state observer \footnote{The word ``observer'' is used in the system-theoretic sense of state estimator.}, access the entire 3-vectors $ \varrho^{0j}$ and $  \varrho^{j0}$ rather than just the sums of their $ ( \lambda_1, \, \lambda_2 ) $ components, then several control designs are possible.
We will not investigate the state estimation issue in detail here, see e.g. \cite{DAlessandro6}. 
We only notice that the system is easily seen to be observable as soon as it is controllable \cite{Albertini3} but that due to the complicated structure of $ {\cal S}$, linear state observers are inadequate (they do not preserve $ \| \bmrho \|$). 
In the heteronuclear case, the estimation of $ \bmrho_\alpha $ and $ \bmrho_\beta $ is an easier task than in the homonuclear one.
In fact, for different spin species, different coils (tuned at the different, well-separated, precession frequencies $ \omega_{o,\alpha } $, $ \omega_{o,\beta} $) can be used so that both pairs $ ( \varrho^{01}, \, \varrho^{02} ) $ and $ ( \varrho^{10}, \, \varrho^{20} ) $ can be available simultaneously from direct measurements.
Since in NMR experiments the initial condition $ \bmrho(0)$ is always known, the simplest way to recover the $ \varrho^{03} $ and $ \varrho^{30} $ components in this case is obviously to numerically integrate the system.

\begin{figure}[h]
\begin{center}
\includegraphics[width=9cm]{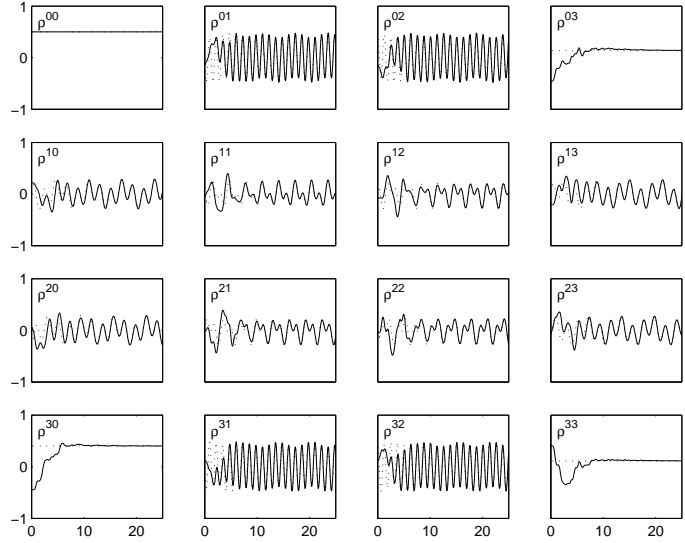}
 \caption{The 16-components of the tensor $ \bmrho_{\rm d} $ (dotted lines) and $ \bmrho$ (solid line) for the orbit tracking problem with feedback from reduced density measurements.}
\label{fig:16-tensor-orb-track-2spin-red-st}
\end{center}
\end{figure}

\begin{figure}[h]
\begin{center}
\includegraphics[width=5cm]{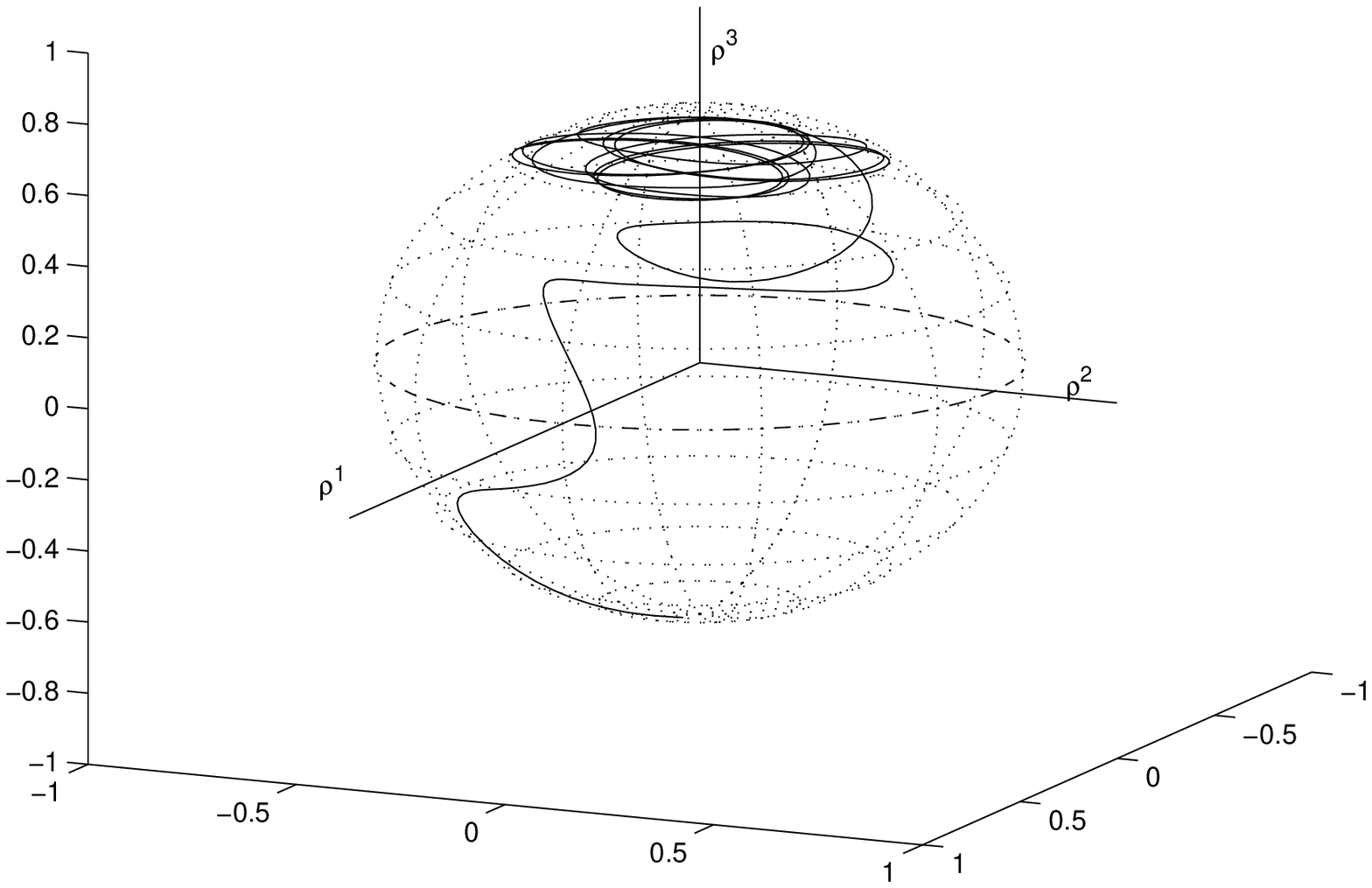}
\includegraphics[width=5cm]{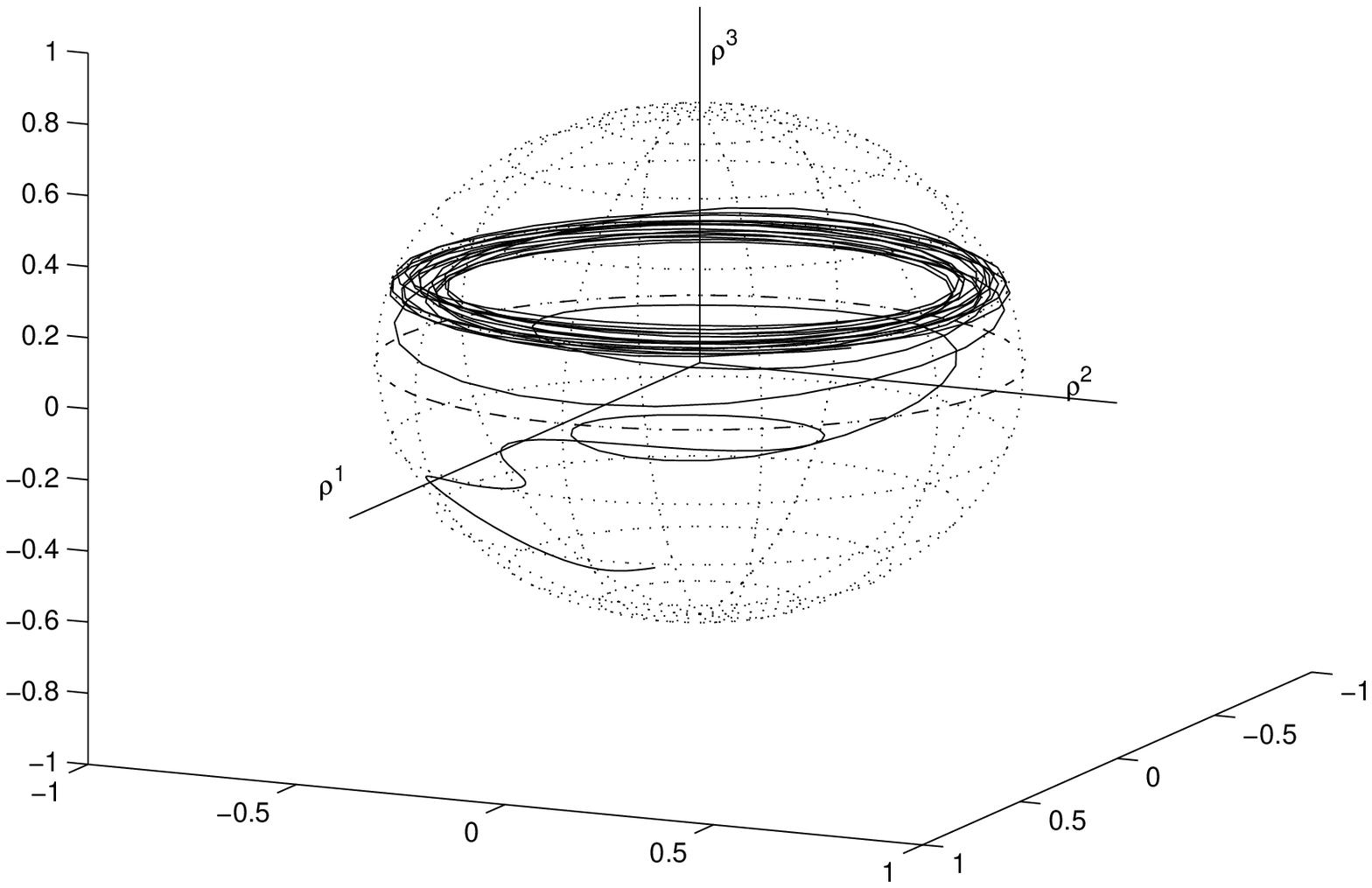}
 \caption{Closed-loop reduced densities $ \bmrho_\alpha (t) $ (left) and $ \bmrho_\beta (t) $ (right) corresponding to Fig.~\ref{fig:16-tensor-orb-track-2spin-red-st}.}
\label{fig:orb-track-red-2spin-red-st}
\end{center}
\end{figure}

Assuming $ \bmrho_\alpha (t) $ and $ \bmrho_\beta(t)$ are available, then one could use $ \bmrho_\alpha (t) \ox \bmrho_\beta(t)$ in place of $ \bmrho(t)$ in the feedback controller of Proposition~\ref{prop:orb-tr-2spin-1c}
\beq
u = k \llangle \bmrho_{\rm d} , \, -i \ad_{H_{\rm c, ns}} \bmrho_\alpha \ox \bmrho_\beta \rrangle .
\label{eq:feedb-orb-tr-red1}
\eeq
The approximation of the true state $\bmrho(t) $ with the product state $ \bmrho_\alpha (t) \ox \bmrho_\beta(t) $ corresponds to disregarding at each time the correlation that is being built by the $J$-coupling $ h^{33}$.
From Remark~\ref{rem:local-feedb-2spin}, this difference is to a large extent negligible.
The results are plotted in Figg.~\ref{fig:16-tensor-orb-track-2spin-red-st}-\ref{fig:orb-track-red-2spin-red-st}: comparing with Figg.~\ref{fig:16-tensor-orb-track-2spin}-\ref{fig:orb-track-red-2spin}, notice how the transient is only slightly worsened.

\section{Design of open loop control via model-based feedback}
\label{sec:model-based-feedb}
\begin{figure}[h!]
\begin{center}
\includegraphics[width=7cm]{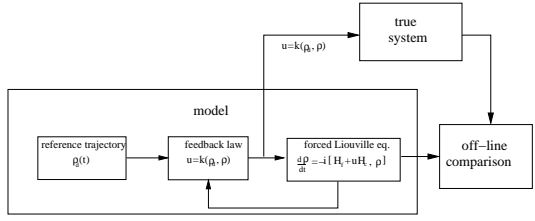}
\caption{Block diagram for the ``model-based feedback'' control design.}
\label{fig:MIT05-feedb-scheme4}
\end{center}
\end{figure}

\begin{figure}[h!]
\begin{center}
\includegraphics[width=3cm]{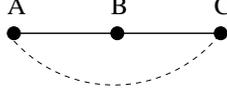}
\caption{Three spin configuration.}
\label{fig:MIT05-3spinc1}
\end{center}
\end{figure}

\begin{figure}[h!]
\begin{center}
\includegraphics[width=4cm]{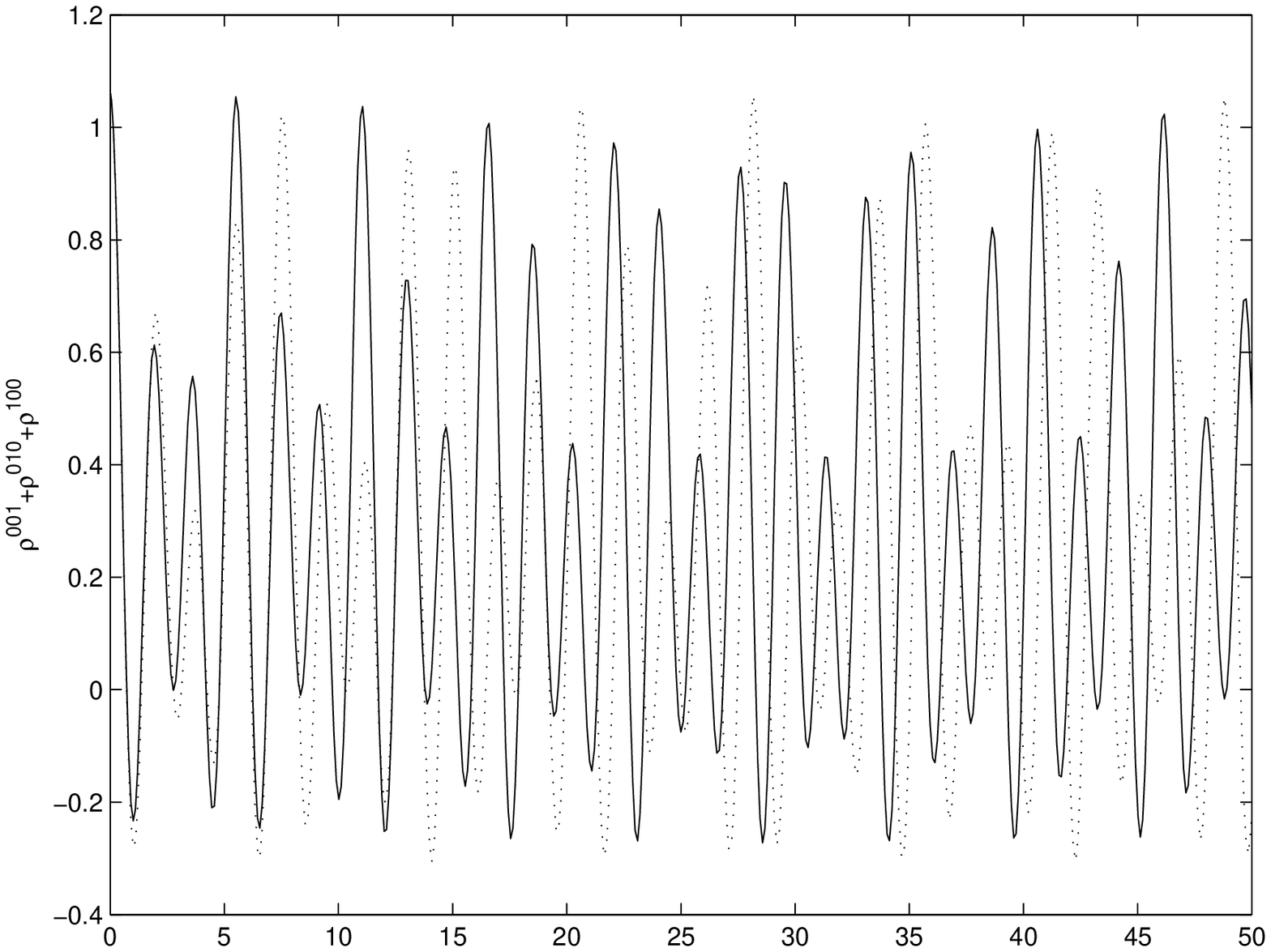}
\includegraphics[width=4cm]{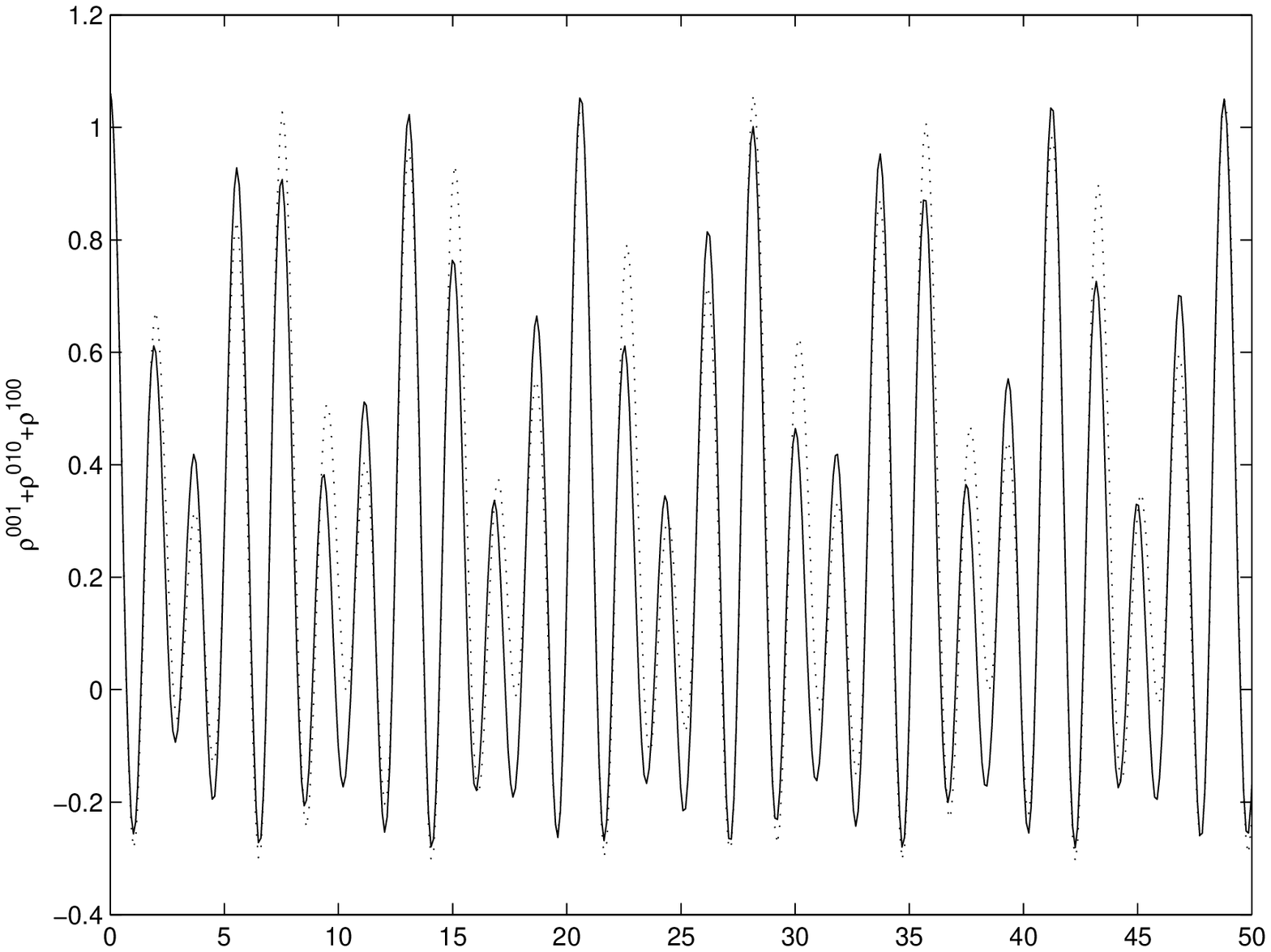}
\caption{Left: Uncontrolled FID response. Right: FID response with the feedback control.}
\label{fig:ens-feedb-ex4-XXX}
\end{center}
\end{figure}

The feedback schemes of the previous Sections have a straightforward application as open-loop model-based state steering methods, see Fig.~\ref{fig:MIT05-feedb-scheme4}.
In this setting, the system and its feedback controller are simulated on a computer.
The time-varying control signal $u(t)$  obtained in this way can then be used in an open-loop fashion to control the true system in the lab.
This strategy is particularly significant for very difficult control tasks, like canceling unwanted couplings in spin systems not admitting selective controls, like is often the case in solid state NMR.
Consider the 3 identical spins configuration of Fig.~\ref{fig:MIT05-3spinc1}.
Each coupling is of dipole-dipole type, with the strength of the AB and BC couplings equal to 1 while for the AC coupling it is $ 1/8 $.
The lack of any chemical shift makes the three spins indistinguishable (``full'' controllability is therefore lost).
The task is to have the three spins behave like a linear chain, canceling the effect of the AC coupling:
\[
\begin{split}
H_{\rm f_{\rm d}} = & -( \Lambda_{110} + \Lambda_{220} - 2  \Lambda_{330}  ) \\
& - ( \Lambda_{011} + \Lambda_{022} - 2  \Lambda_{033}  ) \\
H_{\delta} = & - \frac{1}{8} \left( \Lambda_{101} + \Lambda_{202} - 2  \Lambda_{303} \right)
\end{split}
\]
by means of the control Hamiltonian
\[
H_{\rm c} = u ( \Lambda_{001} + \Lambda_{010} +  \Lambda_{100} ).
\]
Since the purposes is now only to attain a useful control signal $ u$, we can choose the initial condition $ \bmrho(0)= \bmrho_{\rm d}  $. 
Normally in the laboratory one uses $ \bmrho(0) = \bmrho_\alpha (0) \ox \bmrho_\beta (0) \ox \bmrho_\gamma (0) $ where each reduced density is aligned along the $ \lambda_1 $ axis.
Using the feedback scheme of \eqref{eq:feedb-n-spin} the so-called FID (free induction decay) signal $ \varrho^{001} + \varrho^{010}+ \varrho^{100} $ with and without control is shown in Fig.~\ref{fig:ens-feedb-ex4-XXX}.
The same tracking behavior holds for the entire tensor $ \bmrho$. 
Hence the $ H_\delta$ coupling is effectively suppressed on the entire Hilbert space of the system.
Due to the symmetries of the problem, in this case the local precession are totally ininfluent. 
No pulse sequence is known in the NMR literature able to achieve this decoupling, not even if we restrict to the one-dimensional submanifold corresponding to the FID signal and disregard what happens on the rest of the Hilbert space.
This shows the potential of model-based feeback methods even for the usual task of open-loop control design.

\section{Conclusion}

In the classical world, feedback can provide a degree of robustness to all tasks requiring state manipulation or protection which is by far unachievable by open loop control methods.
If one adds that the complexity of the feedback synthesis is much lower than that of open loop design \cite{Sontag1}, it is easy to realize which one control engineers like it better.
The main purpose of this paper is to show that to some extent known feedback methods could be used also in NMR systems.
Unlike other contexts, where ``quantum feedback'' is studied \cite{Doherty2,Hopkins1,Rabitz1,Steck1}, the complete noninvasivity of the bulk measurements makes it theoretically sound -at least in principle- to study the feedback problem for NMR systems in a purely deterministic and unitary framework. 
A possible feedback synthesis follows directly from standard control-theoretic Lyapunov techniques.

\section{Acknowledgments}
The author would like to thank A. Bombrun and H. Kampermann for discussions on the topic of the paper. Part of this work was completed while the author was visiting the Dep. of Nuclear Engineering, MIT. T. F. Havel and D. Cory are gratefully acknowledged.

\bibliographystyle{abbrv}

% \bibliography{/home/altafini/tex/bib/nonhol,/home/altafini/tex/bib/mobman,/home/altafini/tex/bib/quant}

\end{document}